\documentclass[nofootinbib,aps,prd,preprint,
showpacs,amssymb]{revtex4}

\usepackage{amsmath,amsfonts}
\usepackage{wrapfig}
\usepackage{graphicx}
\usepackage{color}
\usepackage{dcolumn}

\allowdisplaybreaks

\newcommand{\be}{\begin{equation}}
\newcommand{\ee}{\end{equation}}
\newcommand{\bea}{\begin{eqnarray}}
\newcommand{\eea}{\end{eqnarray}}

\newcommand{\bs}{\begin{subequations}}
\newcommand{\es}{\end{subequations}}
\newcommand{\pa}{\partial}
\def\no{\nonumber \\}

\begin{document}

\title{Third post-Newtonian accurate generalized quasi-Keplerian 
parametrization for compact binaries in eccentric orbits}

\author{Raoul-Martin Memmesheimer}
\author{Achamveedu Gopakumar} \author{Gerhard  Sch\"afer}

\affiliation{Theoretisch-Physikalisches Institut,
Friedrich-Schiller-Universit\"at Jena, Max-Wien-Platz 1, 07743 Jena, Germany}

\date{\today}

\begin{abstract}
We present Keplerian-type parametrization for the solution of
third post-Newtonian (3PN) accurate equations of motion for two non-spinning
compact objects moving in an eccentric orbit.
The orbital elements of the parametrization are explicitly given 
in terms of the 3PN accurate
conserved orbital energy and angular momentum
in both Arnowitt, Deser, and Misner-type and harmonic coordinates.
Our representation will be required to construct post-Newtonian
accurate `ready to use' search templates for the detection
of gravitational waves from compact
binaries in inspiralling eccentric orbits.
Due to the presence of certain 3PN accurate 
gauge invariant orbital elements, the parametrization should be
useful to analyze the compatibility of 
general relativistic numerical simulations involving compact binaries
with the corresponding post-Newtonian descriptions.
If required, the present parametrization will also be needed to
compute post-Newtonian corrections to the currently employed
`timing formula' for the radio observations of relativistic 
binary pulsars.

\end{abstract}


\maketitle

\section{Introduction}
\label{IntroSec}
   An accurate analytic description for the orbital dynamics 
of compact binaries of comparable masses is highly desirable to probe 
astrophysical scenarios involving strong gravitational fields.
These include inspiralling of compact binaries,
relevant as sources of gravitational radiation for the operational and
proposed interferometric detectors and 
relativistic binary pulsars.

The method of matched filtering, ideally suited to detect gravitational 
radiation emitted by inspiralling compact binaries, requires 
`ready to use' search templates \cite{S91}.
These templates usually  consist of gravitational wave polarizations,
$h_+(t)$ and $ h_{\times}(t)$,
temporally evolving due to the orbital dynamics of the binary.
In the inspiral regime, the orbital dynamics is well described by
the post-Newtonian (PN) approximation to general relativity, which 
allows one to express the equations of motion for a compact binary 
as corrections to Newtonian equations of motion
in powers of $ ( v/c)^2 \sim GM/(c^2 R)$, where $v, M,$ and $R$ are
the velocity, total mass and relative separation of the binary.
Currently, widely employed search templates are 
for compact objects moving in quasi-circular orbits,
where $h_+(t)$ and $h_{\times}(t)$ are known, both in amplitude
and phase evolution, to second post-Newtonian (2PN) order \cite{2PN_hcp}.
However, it is highly desirable to know {\it a priori } 
the temporal phase 
evolution of $h_+(t)$ and $h_{\times}(t)$ at least up to third
post-Newtonian (3PN) order beyond the (Newtonian) quadrupole contribution
\cite{CFPS93}, which in turn requires the knowledge of the conservative binary
dynamics to 3PN order, {\it i.e}, $(v/c)^6$ corrections
to Newtonian equations of motion.
Very recently, after tackling for years many conceptual and computational
issues, the phase evolution of gravitational wave polarizations to
3PN order was achieved, [see \cite{BDFI} and references therein].
Also recently, a formalism to compute `ready to use' search templates for 
compact binaries in inspiralling eccentric orbits was developed and
explicitly implemented at 2.5PN order \cite{DGI}.
Its natural extension to higher PN orders,
which should be relevant for detecting compact binaries in eccentric orbits,
requires a parametric solution to the 3PN accurate equations of motion.

To describe the late stages of binary inspiral and the subsequent merger,
the emphasis is currently being placed on 
numerical relativity, which attempts to
solve the associated set of full Einstein equations 
using supercomputers \cite{NR_rev}.
However, these general relativistic simulations usually do not incorporate
inputs from PN dynamics and are incapable of describing 
few binary orbits even in the inspiral regime.
Recently, a `post-Newtonian diagnostic tool',  
was introduced with the aim of 
extracting some physical information from the
existing general relativistic simulations
\cite{MW02,MW04}.
This `diagnostic tool' requires a solution of 
3PN accurate equations of motion for compact binaries
moving in non-circular orbits.
The tool further introduced 
certain definitions for the eccentricity and semilatus rectum 
parameters in terms of
the orbital angular velocity at the turning points of the orbit,
making many of its estimates gauge dependent.
It is desirable  to characterize a non-circular orbit in post-Newtonian
relativity in terms of gauge invariant quantities and the
parametrization we present here should be useful in that aspect.

  Finally, we note that the high precision radio-wave
observations of binary pulsars employ an accurate relativistic `timing
 formula' \cite{DD86,DT92}  which requires an analytic 
solution to the relativistic  equations of
motion for a compact  binary moving  in an  elliptical orbit
\cite{DD85}.
This `timing formula' is instrumental to  
test both the predictions of general relativity and the 
viability of alternate theories of gravity in strong field situations
\cite{IS03_LR}. 
It should be noted that the long term measurements of the
recently discovered relativistic double pulsar
system J0737-3039 will require the inclusion of higher order 
PN effects in the timing formula \cite{K51004}.

In this paper, we derive an analytic parametric solution to 3PN accurate 
conservative equations of motion for compact binaries 
moving in eccentric orbits.
The orbital representation is given both in 
Arnowitt, Deser, and Misner (ADM)- type and harmonic coordinates.
We employ similar techniques which allowed
Damour and Deruelle to 
obtain a remarkably simple parametrization  
for the solution of 1PN accurate equations of motion for
compact binaries in eccentric orbits
\cite{DD85}. 
Further, our 3PN accurate orbital representation
is structurally quite close to the generalized quasi-Keplerian
representation obtained by Damour, Sch\"afer, and Wex 
for the solution of 2PN accurate orbital motion of compact binaries
in eccentric orbits in ADM gauge \cite{DS88,SW93}. 
We explicitly demonstrate that, to these high PN orders, a slowly 
precessing eccentric orbit can be characterized by certain gauge invariant
quantities. This feature of the parametrization should make it very 
attractive to describe eccentric orbits 
in post-Newtonian relativity.
We note that apart from some misprints, an incomplete
representation in ADM-type coordinates
was obtained earlier \cite{ME01}.

We have the following plan for the paper.
In Sec.\ \ref{Kep_review},
we first review Keplerian parametrization associated with the Newtonian
accurate orbital motion. This will be followed by brief descriptions
about the quasi-Keplerian and the
generalized quasi-Keplerian parametrizations
associated respectively with the 1PN and 2PN accurate orbital motion.
Sec.\ \ref{3PN_ADM} deals with the derivation of 3PN accurate 
orbital representation for compact binaries moving in eccentric orbits
in ADM-type coordinates.
In Sec.\ \ref{3PN_DD}, a similar  parametrization, 
valid in harmonic gauge, is derived.
Finally, in Sec.\ \ref{Disc}, we summarize our results and discuss
the merits and applications of the parametrization.

\section{The Keplerian parametrization and its PN extensions}
\label{Kep_review}

\begin{figure}
\resizebox{16cm}{!}{\includegraphics{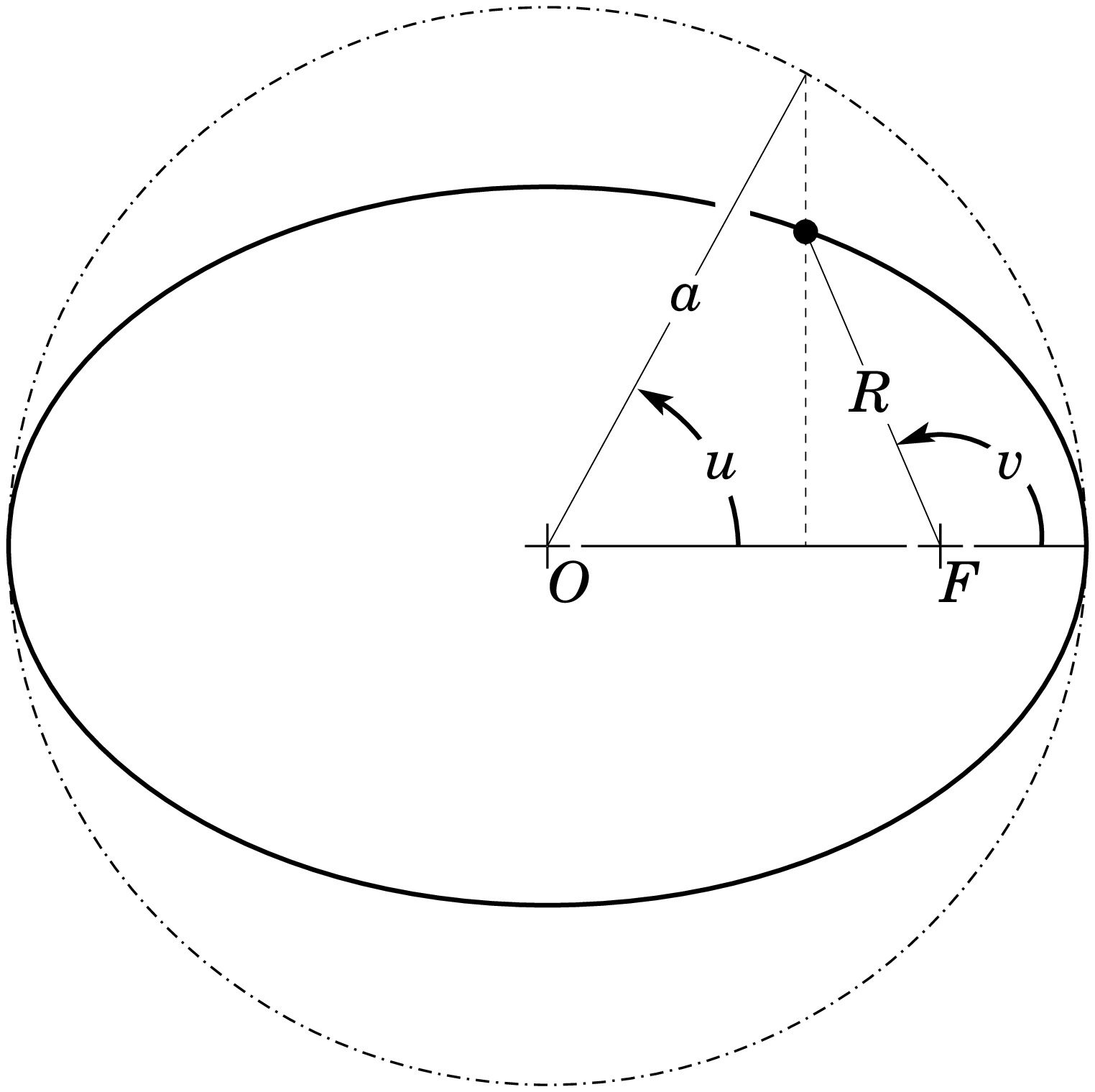}}
\caption {\label{f1}  The geometrical interpretation of 
the eccentric and true anomalies, $u$ and $v$, appearing in the Keplerian 
parametrization. The auxiliary circle of radius $a$
circumscribes the orbital ellipse with semi-major axis $a$.
The points $O$ and $F$ stand for the origin and the focus of the ellipse.
}
\end{figure}

The Keplerian parametrization for Newtonian accurate 
orbital motion of a binary
in eccentric orbit is heavily employed in celestial mechanics \cite{CM_TB}.
In polar coordinates and in the center-of-mass reference frame, the eccentric
motion is parametrized in the following way
\bs
\bea
R &=& a(1-e\cos u )\,,
\\
\phi -\phi_0 &=&v\equiv 2 \arctan \biggl [ \biggl ( \frac{ 1 + e}{ 1 - e}
\biggr )^{1/2} \, \tan \frac{u}{2} \biggr ]\,,
\eea
\es
where $R$ and $\phi$ define the components of the relative separation
vector 
${\bf R} = R ( \cos \phi, \sin \phi, 0)$. 
The semi-major axis and the eccentricity of the orbit are denoted by 
$a$ and $e$ respectively; both are expressible in terms of the Newtonian
conserved (orbital) energy  and angular momentum.
The auxiliary angles $u$ and $v$ are 
called  eccentric and true anomaly respectively.
The geometrical interpretation of these anomalies is 
presented in Fig.\ \ref{f1}.
The explicit time dependence is provided by the Kepler equation, which reads
\be
l \equiv n (t - t_0)  = u - e\,\sin u\,,
\ee
where $l$ is the mean anomaly and $n$ is referred to as the mean motion 
and is given by
$ n = \frac{2\,\pi}{P}$, $P$ being the orbital period.
The quantities $t_0$ and $\phi_0$ are some initial time and 
orbital phase.  The explicit functional dependence of the
orbital elements $a, e$ and 
$n$ is given by
\bs
\bea
a &=& \frac{G\,M}{{(-2\,E)}}\,,\\
e^2 &=& 1 + 2\,E\,h^2\,, \\
n&=& \frac{{{(-2\,E)}}^{3/2}}{GM}\,, 
\eea
\es
where $E$ is the Newtonian orbital energy 
per unit reduced mass $\mu = m_1\,m_2/M$, 
$m_1$ and $m_2$ being the individual masses of the binary and $M= m_1+m_2$.
The reduced angular momentum $h$ is given by $h = \frac{J}{G\,M}$, where
$J$ is the reduced Newtonian orbital angular momentum. 

   For 1PN accurate equations of motion, in harmonic coordinates,
Damour and Deruelle found  the following 
`Keplerian like' parametrization \cite{DD85}, which reads
\bs
\bea
R &=&  a_r(1-e_r \cos u)\,,\\
l \equiv n (t -t_0) &=& u - e_t \sin u\,,\\
\frac{2\,\pi}{\Phi} \,(\phi-\phi_0) &=& v \equiv
2 \arctan \biggl [ \biggl ( \frac{ 1 + e_{\phi}}{ 1 - e_{\phi}}
\biggr )^{1/2} \, \tan \frac{u}{2} \biggr ]\,.
\eea
\es
Note that the three eccentricities $e_r, e_t$ and $e_{\phi}$,
which differ from each other by PN corrections in terms of 
$E$ and the finite mass ratio $\eta = {\mu}/{M}$, were introduced
such that the  parametrization looks `Keplerian' even at 1PN order.
The factor $\frac{2\,\pi}{\Phi}$ 
gives the angle of advance of the periastron
per orbital revolution. Due to these features, in the literature,
the above representation is usually referred to
as the `quasi-Keplerian'
parametrization for 1PN accurate orbital motion.
The parameters $a_r, e_r, e_{\phi}, n $ and $e_t$  are some 1PN
`semi-major' axis, `radial eccentricity', `angular eccentricity',
`mean motion' and `time eccentricity'  respectively.
These orbital elements depend on the 1PN accurate 
conserved energy and angular momentum;
their explicit dependence was derived in \cite{DD85}.

   The above orbital parametrization was extended,
in ADM coordinates, to 2PN order by
Damour, Sch\"afer, and Wex \cite{DS88,SW93}.
The 2PN accurate orbital parametrization has the following form
\bs
\bea
R &=& a_r \left ( 1 -e_r\,\cos u \right )\,,\\
l \equiv n \left ( t - t_0 \right ) &=&
u -e_t\,\sin u +
\left ( \frac{g_{4t}}{c^4} \right )\,(v -u)
 +
  \left ( \frac{f_{4t}}{c^4}  \right )\,\sin v \,,\\
\frac{ 2\,\pi}{ \Phi} \left (\phi - \phi_{0} \right )
 &=& v +
 \left ( \frac{f_{4\phi}}{c^4}  \right )\,\sin 2v
 +
 \left ( \frac{g_{4\phi}}{c^4}  \right )\, \sin 3v\,,
\eea
\es
where $ v = 2 \arctan \biggl [ \biggl ( \frac{ 1 + e_{\phi}}{ 1 - e_{\phi}}
\biggr )^{1/2} \, \tan \frac{u}{2} \biggr ]$.

 The presence of the orbital functions 
$g_{4t}, f_{4t}, g_{4{\phi}}$ and $ f_{4{\phi}}$
indicates that the
structure of the solution is more general than the 
`quasi-Keplerian' one,  prompting to coin
the above parametrization  as the
`generalized quasi-Keplerian' parametrization for  2PN accurate
orbital motion for compact binaries in eccentric orbits.
The explicit expressions for the orbital elements  
$a_r, e_r, n, e_t, \Phi, e_{\phi}$ along with 
$g_{4t}, f_{4t}, g_{4{\phi}}$ and $ f_{4{\phi}} $
in terms of the conserved 2PN accurate
energy and angular momentum as well as the finite mass ratio were derived, in ADM coordinates,
in \cite{SW93}.

   In the next two sections, we will derive a similar orbital 
parametrization for both ADM-type and harmonic gauges, which will 
{\em analytically} describe the solution to the 3PN accurate 
equations of motion for a compact binary moving in an `eccentric' orbit.

\section{The 3PN accurate generalized quasi-Keplerian representation in ADM-type coordinates}
\label{3PN_ADM}

In the Hamiltonian formulation of general relativity, advocated 
by Arnowitt, Deser, and Misner \cite{ADM62}, 
an ordinary Hamiltonian for a compact binary is achievable
only up to 2PN order \cite{OOKH74,DS85}.

At the third post-Newtonian order, 
an higher order Hamiltonian which depends on particle 
positions, conjugate momenta and their derivatives 
was derived by Jaranowski and Sch\"afer \cite{JS98}.
Later, using an higher order contact transformation,
the above  
Hamiltonian was 
transformed into an ordinary Hamiltonian
by Damour, Jaranowski, and Sch\"afer \cite{DJS00a}.
We refer to this ordinary Hamiltonian as the 3PN Hamiltonian 
in ADM-type coordinates.
The two unknown coefficients, which initially appeared in the above 
Hamiltonian, were later fixed \cite{DJS01}.
We display below the fully determined
reduced  ordinary 3PN Hamiltonian, in ADM-type coordinates and
in the center-of-mass frame,
compiled using \cite{DJS00b,DJS01} as
\be 
 \label{H_3}
{\cal{H}}({\bf r},{\hat {\bf p}})
= {\cal{H}}_{0}({\bf r},{\hat{\bf p}})
+ \frac{1}{c^2}{\cal{H}}_{1}({\bf r},{\hat{\bf p}})
+\frac{1}{c^4} {\cal{H}}_{2}({\bf r},{\hat{\bf p}})
+\frac{1}{c^6} {\cal{H}}_{3}({\bf r},{\hat{\bf p}})\,,
\ee                          
where the Newtonian and post-Newtonian contributions are given by
\bs
\bea
{\cal{H}}_{0}({\bf r},{\hat{\bf p}})&=&
\frac{{\hat{\bf p}}^2}{2} - \frac{1}{r},
\\
 {\cal{H}}_{1}({\bf r},{\hat{\bf p}})&=&
 \frac{1}{8}(3\eta-1)\left({\hat{\bf p}}^2\right)^2
- \frac{1}{2}\left[(3+\eta){\hat{\bf p}}^2+\eta({\bf n} \cdot \hat{\bf p})^2\right]\frac{1}{r} + \frac{1}{2r^2},
\\ {\cal{H}}_{2}({\bf r},{\hat{\bf p}})
 &=& 
 \frac{1}{16}\left(1-5\eta+5\eta^2\right)\left({\hat{\bf p}}^2\right)^3
+ 
\frac{1}{8} \bigg[\left(5-20\eta-3\eta^2\right)\left({\hat{\bf p}}^2\right)^2
\no&&
-2\eta^2({\bf n} \cdot \hat{\bf p})^2{\hat{\bf p}}^2-3\eta^2({{\bf n}} \cdot \hat{\bf p})^4 \bigg]\frac{1}{r}
+ \frac{1}{2} \bigg[(5+8\eta){\hat{\bf p}}^2
\no&&
+3\eta({{\bf n}} \cdot \hat{\bf p})^2\bigg]\frac{1}{r^2}  
- \frac{1}{4}(1+3\eta)\frac{1}{r^3},
\\
{ \cal{H}}_{3}({\bf r},{\hat{\bf p}})& =& \frac{1}{128}\left(-5+35\eta-70\eta^2+35\eta^3\right)\left({\hat{\bf p}}^2\right)^4
\nonumber\\&&
+ \frac{1}{16}\bigg[
\left(-7+42\eta-53\eta^2-5\eta^3\right)\left({\hat{\bf p}}^2\right)^3
\no&&
+ (2-3\eta)\eta^2({\bf n} \cdot \hat{\bf p})^2\left({\hat{\bf p}}^2\right)^2
+ 3(1-\eta)\eta^2({\bf n} \cdot \hat{\bf p})^4{\hat{\bf p}}^2 
\no&&
- 5\eta^3({\bf n} \cdot \hat{\bf p})^6
\bigg]\frac{1}{r}
+\bigg[ \frac{1}{16}\left(-27+136\eta+109\eta^2\right)\left({\hat{\bf p}}^2\right)^2 
\no&&
+ \frac{1}{16}(17+30\eta)\eta({\bf n} \cdot \hat{\bf p})^2{\hat{\bf p}}^2 + \frac{1}{12}(5+43\eta)\eta({\bf n} \cdot \hat{\bf p})^4
\bigg]\frac{1}{r^2}
\nonumber\\&&
+\bigg\{ \frac{1}{192} [ -600 + \left(3\pi^2-1340\right)\eta 
- 552\eta^2 ]{\hat{\bf p}}^2 
\no&&
- \frac{1}{64}(340+3\pi^2+112\eta)\eta({\bf n} \cdot \hat{\bf p})^2 
\bigg\}\frac{1}{r^3}
\nonumber\\&&
+ \frac{1}{96}[ 12+ \left(872-63\pi^2\right)\eta] \frac{1}{r^4}\,,
\eea
\es
where $ {\bf r} = {\bf R}/(GM)$, $r= |{\bf r}|$, ${\bf n}={\bf r}/r$ and $ \hat {\bf p} = {\bf P}/\mu$;
${\bf R}$ and ${\bf P}$ are the relative separation vector and
its conjugate momentum vector.

The invariance of ${\cal H}$ under time translation and spatial rotations
leads to the following conserved quantities: 
The 3PN reduced energy $E={\cal H}$ and the reduced
angular momentum $\hat {\bf J} = {\bf r} \times \hat {\bf p}$
of the binary in the center of mass frame.
The conservation of $\hat {\bf J}$ particularly implies that 
the motion is restricted 
to a plane and we may introduce polar coordinates 
such that ${\bf r} = r ( \cos \phi, \sin \phi)$.
The Hamiltonian equations of motion, which govern the relative motion read
\bs
\bea
\dot r &=& \left.{\bf n}\cdot \frac{\pa {\cal H}}{\pa \hat {\bf p}}\right.\,,\\
r^2\, \dot \phi &=& \left|{\bf r} \times \frac{\pa {\cal H}}{\pa \hat {\bf p}}\right|\,,
\label{H_EOMb}
\eea
\label{H_EOM}
\es
where $\dot r = dr/dt, \dot \phi = d \phi/dt $ and $t$ denotes
the coordinate time scaled by $G\,M$.
We introduce $s=1/r$ so that $ \dot r^2 = ( d s/dt)^2/s^4 = \dot s^2/s^4$
and obtain,
using Eqs.\ \eqref{H_EOM}, 3PN accurate expressions for 
$\dot r^2$ and $ d \phi/ds = \dot \phi/\dot s$, in terms of $E, h=|\hat {\bf J}|, \eta$
and $s$, which are displayed in Appendix \ref{AppA}.
We note that the 3PN accurate expressions for $\dot r^2$ and
$\dot \phi$ are seventh degree polynomials in $s$.

To obtain the 3PN accurate orbital parametrization, 
we proceed as follows.
First, we concentrate on the radial motion and 
compute the two nonzero positive roots, having finite limits as
$\frac{1}{c} \rightarrow 0$, of the 3PN accurate expression for
$\dot r^2$.
These two 3PN accurate roots, labeled $s_-$ and $s_+$, 
correspond to the turning points of the radial motion and hence
to the periastron and the apastron of the
post-Newtonian eccentric orbit.
With the help of these two roots, 
we factorize the 3PN accurate expression for $\dot r^2$ 
and obtain the following integral connecting
$t$ and $s$:
\be
t - t_0 = \int_s^{s_-}\frac{A_0+A_1\bar s+A_2\bar s^2
+A_3\bar s^3+A_4\bar s^4+A_5\bar s^5}{\sqrt{(s_--\bar s)(\bar s-s_+)}\,\bar s^2}d \bar s \,.
\label{3PN_t_t0}
\ee
The coefficients $A_i:\,i=1..5$ are some PN accurate 
functions of $E, h$ and $\eta$ which follow directly from 
the expression for $\dot r^2$.
The radial motion is uniquely parametrized by using the ansatz
\be
r = a_r ( 1 -e_r\, \cos u),
\ee
where $a_r$ and $e_r$ are some 3PN accurate semi-major axis and 
radial eccentricity which
may be expressed in terms of $s_-$ and $s_+$ as
\be
a_r = \frac{1}{2} \frac{s_- + s_+}{ s_-\,s_+}\,,
\hspace*{1cm}\,\, e_r = \frac{s_- -s_+}{s_-+\,s_+}\,. 
\ee
With the aid of above relations and 3PN accurate $s_-$ and $s_+$
we compute the 3PN accurate expressions for
$a_r$ and $e_r^2$ in terms of $E, h$ and $\eta$.
We now move on to obtain the 3PN extension to the `Kepler equation'.
For this purpose, we first compute 
the 3PN accurate expression for the radial period $P$
associated with the radial motion.
The expression for $P$ directly follows from Eq.\ (\ref{3PN_t_t0})
and reads
\be
P = 2\, \int^{s_-}_{s_+}\frac{A_0+A_1\bar s+A_2\bar s^2
+A_3\bar s^3+A_4\bar s^4+A_5\bar s^5}{\sqrt{(s_--\bar s)(\bar s-s_+)}\,\bar s^2}d \bar s \,.
\label{3PN_T}
\ee
Using Eq.\ (\ref{3PN_t_t0}), we express the mean anomaly 
$ l \equiv n(t -t_0) 
= \frac{2\,\pi}{P} (t - t_0)$
in terms of the eccentric anomaly $u$.
We introduce an auxiliary variable 
$\tilde  v = 2 \arctan 
\bigg[ {\left( \frac{ 1 + e_{r}}{ 1 - e_{r}}\right)}^{1/2} 
\, \tan \frac{u}{2} \bigg]$ and with the aid of 
certain trigonometric relations
involving $\tilde v$, displayed in Appendix \ref{AppB}, we obtain the
following temporary parametrization for the mean anomaly $l $
which has the following general structure 
\bea
l \equiv n\,(t - t_0) &=& 
u+ \kappa_0 \sin u+
 \frac{\kappa_1}{c^2} (\Tilde v-u)
 \no &&
\quad +\frac{\kappa_2}{c^4} \sin \Tilde v+\frac{\kappa_3}{c^6} \sin 2 \Tilde v
+\frac{\kappa_4}{c^6} \sin 3 \Tilde v\,.
\label{temp_t}
\eea 
The coefficients $\kappa_i$ are some PN accurate 
functions of $E, h$ and $\eta$.
Since these expressions are lengthy and are only required 
temporarily, we do not list them in the paper.

   Let us now turn our attention to the parametrization of the angular part,
which is also required to obtain the final representation
for the PN accurate `Kepler equation'. 
Using the 3PN accurate expression for $\dot \phi$ and the factorized
3PN accurate expression for $\dot s$, we compute
$\frac{d \phi}{d s} = \dot \phi / \dot s$ 
and  readily obtain the following relation between $\phi$ and
$s$ 
\be
\phi -\phi_0 =\int_s^{s_-}
\frac{B_0+B_1\bar s+B_2\bar s^2+B_3\bar s^3+B_4\bar s^4+B_5\bar s^5}{\sqrt{(s_--\bar s)(\bar s-s_+)}}d \bar s\,,
\label{phi_3PN}
\ee                 
where the coefficients $B_i : i=1..5$ are some 
PN functions of $E, h $ and $\eta $ as
presented in  Appendix \ref{AppA}.                                                                                                                              
Next step 
involves the computation of
the advance of periastron during the radial period $P$. 
This quantity, denoted by 
$\Phi$, is obtained by evaluating the
integral
on the right hand side of
Eq.\ (\ref{phi_3PN}) between $s_+$ and $s_-$,
\be
\Phi = 2\, \int^{s_-}_{s_+}
\frac{B_0+B_1\bar s+B_2\bar s^2+B_3\bar s^3+B_4\bar s^4+B_5\bar s^5}{\sqrt{(s_--\bar s)(\bar s-s_+)}}d \bar s\,.
\label{Phi_3PN}
\ee

To derive a temporary parametrization of the angular motion, 
using Eqs.\ (\ref{phi_3PN}) and (\ref{Phi_3PN}), 
we first compute $\frac{2\,\pi}{\Phi}\,( \phi -\phi_0)$.
Employing certain trigonometric
relations, given in Appendix \ref{AppB}, 
we arrive at a temporary 3PN accurate parametrization
for the angular motion 
\bea
\frac{2\,\pi}{\Phi}\,( \phi -\phi_0) &=& \Tilde v
+\frac{\lambda_1}{c^2}\sin \Tilde v
+
\frac{\lambda_2}{c^4}\sin 2 \Tilde v
\no
&&
\quad
+
\frac{\lambda_3}{c^4} \sin 3\Tilde v
+
\frac{\lambda_4}{c^6} \sin 4\Tilde v
+
\frac{\lambda_5}{c^6} \sin 5\Tilde v\,,
\label{temp_phi}
\eea  
where $\lambda_i$ are some PN accurate functions, expressible
in terms of $E, h$ and $\eta$.

  To obtain the final parametrization for $l$ and $\phi$ equations, 
we introduce $v=2\,\arctan\left[\left(\frac{1+e_{\phi}}{1-e_{\phi}}\right)^{1/2} \tan \frac{u}{2}\right]$, where $e_{\phi}$ 
differs from $e_r$ by yet to be determined 
PN corrections.
It is straightforward to express $\tilde v$ in terms of $ v$, 
some trigonometric functions of $v$ and the PN corrections relating
$e_{\phi}$ and $e_r$, as shown schematically in Eq.\ (\ref{tildevofv}).
We systematically introduce $v$ in Eq.\ (\ref{temp_phi}) such that the
coefficient of $\sin v$ is zero up to the third PN order.
This uniquely determines the PN corrections defining $e_{\phi}$ and
consistently generalizes the 
imposed requirement that
the expression for $\frac{2\,\pi}{\Phi}\,( \phi -\phi_0)$
at 1PN order has a `Keplerian' structure.
However, we find that
at 2PN and 3PN orders, the $v$ contribution will be 
supplemented by other trigonometric functions of $v$.
With the introduction of $v$, the 3PN accurate expression for 
$\frac{2\,\pi}{\Phi}\,( \phi -\phi_0)$, given by Eq.\ (\ref{temp_phi}),
becomes
\bea
\frac{2\,\pi}{\Phi}\,( \phi -\phi_0 )&=& v + 
\left ( \frac{f_{4\phi}}{c^4} + \frac{f_{6\phi}}{c^6} \right )\,\sin 2v
+
\left ( \frac{g_{4\phi}}{c^4} + \frac{g_{6\phi}}{c^6} \right )\, \sin 3v
\no
&&
+ \frac{i_{6\phi}}{c^6}\, \sin 4v
+ \frac{h_{6\phi}}{c^6}\, \sin 5v \,,
\label{fin_phi}
\eea   
where $ v=2\,\arctan\left[\left(\frac{1+e_{\phi}}{1-e_{\phi}}\right)^{1/2} \tan \frac{u}{2}\right]$.
The explict 3PN accurate 
expressions for $\Phi, e_{\phi}$ and the orbital functions will be displayed below.

  Let us now go back to Eq.\ (\ref{temp_t}), giving the
parametrization to $l$ in terms of $u, e_t$ and $\tilde v(u)$.
Using the 3PN accurate relation connecting $\tilde v $ and $v$, we rewrite
that equation and  obtain
the following 3PN extension to 
the `Kepler equation':
\bea
l \equiv n\,(t -t_0) &=& u -e_t\,\sin u + 
\left ( \frac{g_{4t}}{c^4} + \frac{g_{6t}}{c^6} \right )\,(v -u)
 \no
&&
 + 
\left ( \frac{f_{4t}}{c^4} + \frac{f_{6t}}{c^6} \right )\,\sin v
+ \frac{i_{6t}}{c^6} \, \sin 2\, v 
+ \frac{h_{6t}}{c^6} \, \sin 3\, v\,.
\label{K_Eqn_3PN}
\eea 
The 3PN accurate expressions for $n, e_t$ and the orbital functions 
appearing in the above `Kepler equation' will be displayed below.

Finally, we display, in its entirety, the third post-Newtonian
accurate 
generalized quasi-Keplerian parametrization for a compact binary
moving in an eccentric orbit in ADM-type coordinates 
\bs 
\label{e:FinalParamADM}
\bea
r &=& a_r \left ( 1 -e_r\,\cos u \right )\,,\\
l \equiv n \left ( t - t_0 \right ) &=&
u -e_t\,\sin u +
\left ( \frac{g_{4t}}{c^4} + \frac{g_{6t}}{c^6} \right )\,(v -u)
 \no
 &&
  +
  \left ( \frac{f_{4t}}{c^4} + \frac{f_{6t}}{c^6} \right )\,\sin v
  + \frac{i_{6t}}{c^6} \, \sin 2\, v
  + \frac{h_{6t}}{c^6} \, \sin 3\, v \,,\\
\frac{ 2\,\pi}{ \Phi} \left (\phi - \phi_{0} \right )
&=& v +
\left ( \frac{f_{4\phi}}{c^4} + \frac{f_{6\phi}}{c^6} \right )\,\sin 2v
+
\left ( \frac{g_{4\phi}}{c^4} + \frac{g_{6\phi}}{c^6} \right )\, \sin 3v
\no
&&
+ \frac{i_{6\phi}}{c^6}\, \sin 4v
+ \frac{h_{6\phi}}{c^6}\, \sin 5v \,,
\eea
\es
where $ v = 2\arctan \biggl [ \biggl ( \frac{ 1 + e_{\phi}}{ 1 - e_{\phi}}
\biggr )^{1/2} \, \tan \frac{u}{2} \biggr ]$.
The 3PN accurate expressions for the orbital elements $a_r, e_r^2,n, e_t^2 , \Phi,$
and $e_{\phi}^2$ and the post-Newtonian orbital
functions 
$  g_{4t},  g_{6t}, f_{4t}, f_{6t}, i_{6t}, h_{6t},
 f_{4\phi}, f_{6\phi}, g_{4\phi}, g_{6\phi},   i_{6\phi}, $
 and $h_{6\phi}$, in terms of $E, h$ and $\eta$ 
read
\bs
\label{e:CoeffKP}
\bea
a_r &=&  \frac{1}{{(-2\,E)}}\bigg\{ 1+\frac{ (-2\, E )}{4\,c^2} \left( -7+\eta \right) +
\frac{{{ (-2\, E) }}^{2}}{16 c^4}\,\bigg[ 
(1+10\,\eta+{\eta}^{2})
 \no&&
 +\frac {1}{(-2\,E\,h^2)}
(  -68+44\,\eta)
\bigg] 
+{\frac {{{ (-2\,E) }}^{3}}{192\,c^6}}\, 
\biggl [ 
3-9\,\eta-6\,{\eta}^{2}
\no&&
+3\,{\eta}^{3}+\frac{1}{(-2\,E\,h^2)}
\biggl (
864+ \left( -3\,{\pi}^{2}-2212 \right) \eta+432\,{\eta}^
{2}\biggr)
\no 
&&
+
\frac{1}
{ (-2\,E\, h^2)^2 }
\biggl (
-6432+ \left( 13488-240\,{\pi}^{2} \right) \eta
-768\,{\eta}^{2}\biggr )
\biggr ]   \bigg\}\,,
\\
{e_{{r}}}^{2} &=&
1 + 2\,E\,h^2 + 
\frac{(-2\,E)}{4\,c^2}
\biggl \{ 24 -4
\,\eta+5\,\left(-3+ \eta \right) {(-2\,E\,h^2)} 
\biggr \} 
\no
&&  
+ \frac{ (-2\,E)^2}{8\,c^4}
\biggl \{
52+2\,\eta+2\,{\eta}^{2}
-\left( 80-55\,\eta+4\,{\eta}^{2} \right) {(-2\,E\,h^2)}
\no&&
-\frac {8}{ (-2\,E\,h^2)}
\left ( -17+11\, \eta \right )
\biggr \}
+ \frac{ (-2\,E)^3 }{ 192\,c^6} \biggl \{
-768-6\,\eta\,{\pi}^{2}
\no&&
-344\,\eta-216\,{\eta}^{2} 
+ 3(-2\,E\, h^2)\,\bigg(
-1488+1556\,\eta
-319\,{\eta}^{2}
\no&&
+4\,{\eta}^{3}
\bigg) \,
-\frac{4}{ (-2\,E\, h^2)}\,\bigg(
588-8212\,\eta+177\,\eta
\,{\pi}^{2}+480\,{\eta}^{2}\bigg) 
\no
&&
+\frac{192}{(-2\,E\,h^2)^2}
\biggl (
134-281
\,\eta+5\,\eta\,{\pi}^{2}+16\,{\eta}^{2}
\biggr )  
\biggr \}\,, 
\\
n&=&{{(-2\,E)}}^{3/2} \bigg\{ 1+{\frac {{(-2\,E)}} {8\,{c}^{2}}}\, 
\left( -15+\eta \right)
+{\frac {{{(-2\,E)}}^{2} }{128{c}^{4}}} 
\biggl [ 555 
+30\,\eta
\no&&
+11\,{ \eta}^{2}
+ \frac{192}{ \sqrt{(-2\,E\,h^2)}}
\left( -5+2\,\eta \right ) 
\biggr ]
+
{\frac {{{(-2\,E)}}^{3}}{3072\,{c}^{6}}}
\biggl [  -29385 
\no&&
-4995\,\eta-315\,{\eta}^{2}+135
\,{\eta}^{3}
-
\frac {16}{( -2\,E\,h^2)^{3/2}}
\bigg(
10080+123\,\eta\,{\pi}^{2}
\no&&
-13952\,\eta+1440\,{\eta}^{2}\bigg)   
+ \frac{5760}{ \sqrt{ (-2\,E\,h^2)}}
\left(17 -9\,\eta+2\,{\eta}^{2} \right )
\biggr ] 
\bigg\} \,, 
\\    
{\it e_t}^2 &=&1+{2\,E}\,{h}^{2}+ 
{\frac {{(-2\,E)}}{4\,{c}^{2}}}\, {\bigg\{ -8+8\,
\eta
- \left( -17+7\,\eta \right) {(-2\,E\, h^2)} \bigg\} }  
\no&&
+   
\frac{{{(-2\,E)}}^{2}}{8\,{c}^{4}} \bigg\{8+4\,\eta +20\,{\eta}^{2}
-  {(-2\,E\,h^2)}( 112-47\,\eta
+16\,{\eta}^{2} )
\no&&
-24\,\sqrt{(-2\,E\,h^2)}\, \left( -5+2\,\eta \right) 
+\frac{4}{(-2\,E\,h^2)} \left( 17 - 11\,\eta \right )
\no&&             
-\frac{24}{ \sqrt{ (-2\,E\, h^2)}}
\, \left ( 5 -2\,\eta \right )
\bigg\}
\no &&
+{
\frac {{{(-2\,E)}}^{3}}{192\,c^{6}}}
\bigg\{ 24\, \left( -2+5\,\eta \right) 
 \left(-23+10\,\eta+ 4\,{\eta}^{2} \right) 
-15\,\biggl (-528
\no&&
+200\,\eta-77\,{\eta}^{2}
+ 24\,{
\eta}^{3} \biggr ) {(-2\,E\, h^2)}
-72\,( 265-193\,\eta
\no&&
+46\,{\eta}^{2} ) \sqrt {{(-2\,E\, h^2)}}
- \frac{2}{(-2\,E\,h^2)}
 \bigg(
6732 +117\,\eta\,{\pi }^{2}-12508\,\eta
\no&&
+2004\,{\eta}^{2}\bigg)
+ \frac{2}{\sqrt{ (-2\,E\,h^2)}}
\bigg(
16380-19964\,\eta+123\,\eta\,{\pi }^{2}
\no&&
+3240\,{\eta}^{2}
\bigg)
-\frac{2}{ (-2\,E\,h^2)^{3/2} }
\bigg( 10080+123\,\eta\,{
\pi }^{2}-13952\,\eta
\no&&
+1440\,{\eta}^{2}
\bigg)
+ \frac{96}{ (-2\,E\,h^2)^2}
\bigg(
134 -281\,\eta+5\,\eta\,{\pi }^{2}+16\,{\eta}^{2}
\bigg)
 \bigg\}\,,
\\
g_{{4\,t}} &=&\frac{3\,(-2\,E)^{2}}{2}\,\biggl \{ \frac{5 -2\,\eta }{ \sqrt{ (-2\,E\,h^2)}}
 \biggr \}\,, 
\\
g_{{6\,t}} &=&{\frac {{{(-2\,E)}}^{3}}{192}}
\biggl \{
\frac{1}{( -2\,E\,h^2)^{3/2} }
\bigg(
10080+123\,\eta\,{\pi }^{2}-13952\,\eta
\no&&
+1440\,{\eta}^{2}
\bigg)
+ \frac{1}{\sqrt{(-2\,E\,h^2)}}
\left ( -3420
+1980\,\eta-648\,{\eta}^{2}
\right )
\biggr \}\,,
\\ 
f_{{4\,t}} &=& -\frac{1}{8}\,
\frac{ (-2\,E)^2}{ \sqrt{ (-2\,E\,h^2)}}
\biggl \{ (4 + \eta)\,\eta \, \sqrt{(1 +2\,E\,h^2)}
\biggr \}\,,  
\\
f_{{6\,t}}&=&{\frac {{{(-2\,E)}}^{3}}{192}}
\bigg\{
\frac{1}{(-2\,E\,h^2)^{3/2} } \,
\frac{1}{ \sqrt{1 +2\,E\,h^2 }}
\bigg(
1728-4148\, \eta +3\,\eta\,{\pi }^{2}
\no&&
+600\,{\eta}^{2}+33\,{\eta}^{3}\bigg)
+3\,
\frac{\sqrt{(-2\,E\,h^2)}}{ \sqrt{ ( 1+2\,E\,h^2)}}
\eta\, \left(-64-4\,\eta
+ 23\,{\eta}^{2} \right)
\no&& 
+ \frac{1}{ \sqrt{ ( -2\,E\,h^2)\,(1 +2\,E\,h^2 )}}
\biggl ( 
-1728
+
4232\,\eta-3\,\eta\,{\pi }^{2}
\no&&
-627\,{\eta}^{2}-105\,{\eta}^{3}
\biggr )
\bigg\}\,, 
\\
i_{{6\,t}} &=&\frac{{{(-2\,E)}}^{3}}{32}\,\eta
\biggl \{ 
\frac{(1 +2\,E\,h^2)}{ (-2\,E\,h^2)^{3/2} }
\left(23+12\,\eta+ 6\,{\eta}^{2} \right) 
\biggr \}\,,  
\\
h_{{6\,t}} &=&
{\frac {13\,{{(-2\,E)}}^{3}}{192}}
\eta^3
\biggl ( 
\frac{  1 + 2\,E\,h^2 }{ -2\,E\,h^2}  
\biggr )^{3/2}\,, 
\\
\Phi&=&2\,\pi \, \bigg\{ 1+{\frac {3}{{c}^{2}{h}^{2}}}+
\frac{{{(-2\,E)}}^{2}}{4\,{c}^{4}}
\biggl [  \frac{3}{(-2\,E\,h^2)} \left ( -5+2\,\eta \right )
\no&&
+ \frac{15}{(-2\,E\,h^2)^2} \left ( 7 -2\,\eta \right )
\biggr ]  
+{\frac {\,{{(-2\,E)}}^{3}}{128\,{c}^{6}}}
\biggl [  
 \frac{24}{ (-2\,E\,h^2)}
( 5 -5\eta 
\no&&
+ 4\eta^2)
- \frac{1}{ (-2\,E\,h^2)^2}
\biggl ( 10080
 -13952\,\eta+123 \,\eta\,{\pi }^{2}+1440\,{\eta}^{2}
\biggr )
\no&&
+ \frac{5}{(-2\,E\,h^2)^3}
\biggl (7392-8000\,\eta+  123\,\eta\,{\pi }^{2}
+ 336\,{\eta}^{2} \biggr  )
\biggr ] 
 \bigg\}\,, 
\\
f_{{4\,\phi}} &=&
\frac{{{(-2\,E)}}^{2}}{8}
\,\frac{( 1+2\,E\,h^2)}{(-2\,E\,h^2)^2}\,
\eta \, (1 -3\,\eta)\,,
\\
f_{{6\,\phi}}&=&{\frac {{{(-2\,E)}}^{3}}{256}}
\bigg\{
\frac{4\,\eta}{(-2\,E\,h^2)}
\left( -11-40\,\eta+24\, {\eta}^{2} \right)
\no&& 
+ \frac{1}{{(-2\,E\,h^2)}^2} \biggl (
-256
+1192\,\eta-49\,\eta\,{\pi }^{2}
+336\,{\eta}^ {2}
-80\,{\eta}^{3}
\biggr )
\no&&
+ \frac{1}{(-2\,E\,h^2)^3}
\biggl ( 
256+49\,\eta\,{\pi }^{2}-1076\,\eta-384\,{\eta}^{2}-40\,{\eta}^{3}
\biggr )
\bigg\}\,,
\\
g_{{4\,\phi}} &=&
-{\frac {3{{(-2\,E)}}^{2}}{32}}
\frac{\,\eta^2\,}{(-2\,E\,h^2)^2}
 ( 1 +2\,E\,h^2)^{3/2}\,,
\\
g_{{6\,\phi}}&=&
\frac{ (-2\,E)^3}{768}\, \sqrt{(1 +2\,E\,h^2)}\,
\bigg\{
-\frac{3}{ (-2\,E\,h^2)}\,\eta^2\, \left( 9-26 \,\eta \right )
\no   &&
- \frac{1}{(-2\,E\,h^2)^2} \,\eta
\biggl ( 220+3\,{\pi }^
{2}
+312\,\eta+150\,{\eta}^{2} \biggr )
\no&&
+ \frac{1}{(-2\,E\,h^2)^3}\,\eta
 \left( 220+3\,{\pi }^{2}+96\,\eta+45\,{\eta}^{2}
\right )
\bigg\}\,,
\\
i_{{6\,\phi}} &=&{\frac {{{(-2\,E)}}^{3}}{128}}
\,\frac{{(1 +2\,E\,h^2)}^2}{(-2\,E\,h^2)^3}\,\eta 
\left( 5+28\,\eta+10\,{\eta}^{2} \right)\,,
\\
h_{{6\,\phi}} &=&
\frac{5\, (-2\,E)^3}{256}\, \frac{\eta^3}{ (-2\,E\,h^2)^3}
\, (1 +2\,E\,h^2)^{5/2} \,,
\\
{e_{{\phi}}}^{2}&=&
1 + 2\,E\,h^2 +
{\frac {{(-2\,E)} }{4\,{c}^{2}}} \bigg\{ 
24+ \left( -15+\eta \right) {(-2\,E\,h^2)} \bigg\}   
\no&&
+\frac{{{(-2\,E)}}^{2} }{16\,{c}^{4}}
\bigg\{ -32+ 176\,\eta+18\,{\eta}^{2} 
- {(-2\,E\,h^2)}( 160-30
\,\eta
\no&&
+3\,{\eta}^{2} )
+ \frac{1}{(-2\,E\,h^2)}\,
\left ( 408 - 232\,\eta - 15\,{ \eta}^{2} \right )
\no
&&
+{\frac {{{ (-2\,E)}}^{3}}{384\,{c}^{6}}}
\bigg\{ -16032+2764\,
\eta+3\,\eta\,{\pi}^{2}+4536\,{\eta}^{2}+234\,{\eta}^{3}
\no&&
-36\, \biggl ( 248
-80\,\eta
+13 \,{\eta}^{2}+{\eta}^{3} \biggr )\, {(-2\,E\,h^2)}
-
\frac{6}{(-2\,E\,h^2)}
\bigg( 2456
\no&&
-26860\,\eta+581\,\eta\,{\pi}^{2}+2689\,{\eta}^{2}+10\,{\eta}^{3}
\bigg)
+ \frac{3}{(-2\,E\,h^2)^2}
\biggl ( 
27776
\no&&
-65436\,\eta+1325\,\eta\,{\pi}^{2}+3440\,{\eta}^{2}-70\,{\eta}^{3}
\biggr )\biggr \}\,.
\eea
\es 
The three eccentricities
$e_r, e_t$ and $e_\phi$, which differ from each other
at PN orders, are related by
\bs
\bea
{e_t}&=&{e_r} \bigg\{1+ {\frac {{(-2E)}}{2{c}^
{2}}}(-8+3\eta)+  {\frac {{{(-2E)}}^{2}}{8{c}
^{4}}} {\frac {1}{{(-2E{h}^{2})}}}
\bigg[-34+22\eta
\no&&
+
\left(
-60+24\eta \right) \sqrt {{(-2Eh^2)}}
+
 \left(72 -33\eta+12{\eta}^{2} \right){(-2E {h}^{2})}
\bigg]
\no&&
+{\frac {{{(-2E)}}^{3}}{192{c}^{6}}}
{\frac {1}{{{(-2E{h}^{2})}}^{2} }}
\bigg[
-6432+13488\eta-240\eta{\pi}^{2}
\no&&
-768{\eta}^{2}
+
\bigg( -10080+13952\eta-123\eta{\pi}^{2}
\no&&
-1440{\eta}^{2} \bigg)
\sqrt {{(-2Eh^2)}}
+
(2700-4420\eta-3\eta{\pi}^{2}
\no&&
+1092{\eta}^{2} ) {(-2E{h}^{2})}
+
( 9180-6444\eta+1512{
\eta}^{2}){{(-2E {h}^{2})}}^{3/2}
\no&&
+
\bigg( -3840+1284\eta-672{\eta}^{2}
+240{\eta
}^{3} \bigg) {{(-2E{h}^{2})}}^{2}
\bigg] \bigg\}
\,,\\
{ e_\phi}&=&{ e_r} \bigg\{ 1+  {\frac {{ (-2E)}}{2{c}^
{2}}} \eta+  {\frac {{{ (-2E)}}^{2}}{32{c}^{4}}
}  {\frac {1}{{ (-2E{h}^{2})}}} \bigg[136-56\eta-15{
\eta}^{2}
\no&&
+\eta( 20+11{\eta}) { (-2E{h}^{2})}\bigg]+  {\frac {{{ (-2E)}}^{3}}{768{c}^{6}
}}   {\frac {1}{{{ (-2E{h}^{2})}}^{2}}} \bigg[
31872
\no&&
-88404\eta+2055 \eta{\pi}^{2}+4176{\eta}^{2}-210{\eta}^{3}+
\bigg(2256
\no&&
+10228\eta -15\eta{\pi}^{2}
-2406{\eta}^{2}-450{\eta}^{3} \bigg) { (-2E{h}^{2})}+
 6\eta\, ( 136
\no&&
+34{\eta}+31{\eta}^{2} ){{
(-2E {h}^{2})}}^{2}
\bigg] \bigg\}\,.
\eea
\es
These relations allow one to choose a specific eccentricity,
while describing a PN accurate non-circular orbit.

  It is highly desirable that such a detailed parametrization and lengthy 
expressions for  the PN accurate orbital elements and functions 
should be subjected to possible consistency checks.
We have devised and performed the following consistency check for 
our computation.
Note that the expressions for $a_r$ and $e_r^2$ were obtained 
from the PN accurate roots $s_-$ and $s_+$,  representing 
the turning points of the radial
motion. This prompted us to express the expressions for $\dot r^2 $ and
$\dot \phi^2$, derived using the Hamiltonian equations of motion and given
by Eqs.\ (\ref{H_EOM}), in terms of $E, h, \eta$ and $(1-e_r\,\cos(u))$.
We then compared the above expression for $\dot r^2$ with the one derived
using the parametrization, namely
$\dot r^2 = \left ( \frac{d r}{du}\, \frac{du}{dt} \right )^2$.
This expression for $\dot r^2$, after some lengthy algebra, is
found to be in total agreement
with $\dot r^2$, computed using Hamiltonian equations of motion to the
third post-Newtonian order.
We note that the above 
computation fully checked both the structure and parameters 
of both the radial and temporal part of the generalized quasi-Keplerian
representation.
We performed a similar check on the angular part by noting that
${\dot \phi}^2 = {\left ( \frac{d \phi}{dv}\, \frac{d v}{du}\, \frac{du}{dt}
\right )}^2 $.
The expressions for ${\dot \phi}^2$, computed using 
the above  relation and via the Hamiltonian equations of motion,
were also found to be in total agreement
to 3PN order.
These two computations provided us with powerful checks on our 
3PN accurate generalized quasi-Keplerian parametrization.

Finally, we observe that to the 2PN order, our results are in agreement with
results available in \cite{DS88,SW93}. The above parametrization
is also consistent with results
given in \cite{ME01}, modulo typographical errors and omissions.
In the next section, we obtain a similar parametrization in harmonic
gauge.

\section{The 3PN accurate generalized quasi-Keplerian 
parametrization in harmonic coordinates}
\label{3PN_DD}
Incidentally, it was in harmonic coordinates that the quasi-Keplerian 
parametrization for compact binaries of arbitrary mass ratio moving in 
eccentric orbits was first realized \cite{DD85}.
The above computation employed 1PN accurate expressions for the conserved
orbital energy and angular momentum in terms of the dynamical variables 
of the binary to derive 
1PN accurate expressions, in harmonic coordinates, for 
$\dot r^2$ and $\dot \phi^2$.
However at 2PN order, the parametrization was achieved only in 
ADM gauge \cite{DS88,SW93}, though the description of the binary
dynamics, in these two gauges,  begins to differ 
at 2PN order \cite{DS85}.

        In this section, we will compute 3PN accurate generalized 
quasi-Keplerian parametrization, in harmonic coordinates, for 
compact binaries in eccentric orbits. 
To begin our computation,
we will require 3PN accurate expressions for the conserved
orbital energy and angular momentum in harmonic coordinates 
and in the center-of-mass frame.
These quantities, written in terms of the dynamical variables of the binary,
are required to derive, by iteration, 
3PN accurate expressions for 
$\dot r^2$ and $\dot \phi^2$ in terms of $E,h$ and $r$.
The above mentioned 3PN accurate conserved orbital 
energy and angular momentum, in harmonic coordinates, 
are available in \cite{BI03}
[For a 
comprehensive review of the post-Newtonian computations, 
especially in harmonic coordinates,
see \cite{LB_lr} ]. However, we will not be able to employ 
directly these conserved quantities in our computation for 
$\dot r^2$ and $\dot \phi^2$ for the following two reasons.
First, at 3PN order, these expressions  contain logarithmic terms involving the
radial separation $r$. These logarithmic terms will not allow us to
obtain 3PN accurate expressions for $\dot r^2$ and $\dot \phi^2$ as 
polynomials in $s=1/r$, preventing the determination of 3PN accurate 
roots of $ \dot r^2$ and hence the parametrization 
in harmonic coordinates. However, in 
the near-zone of a gravitating system, the harmonic gauge conditions
do not fix the coordinate system  uniquely, allowing 
various harmonic coordinates to be employed at 3PN order
\cite{BF00,BF01,IF03,I04}.
Further, it is possible to remove these logarithmic terms using
a 3PN accurate coordinate transformation that still respects the
harmonic gauge condition \cite{BF01}.
The problematic logarithmic terms appearing in the 3PN
accurate expressions for the orbital 
energy and angular momentum, as available in \cite{BI03},
were removed using the coordinate transformation given
in \cite{BF01}.
Secondly, the expression for 3PN accurate orbital energy
appearing in \cite{BI03} involves an undetermined
parameter, the so called
regularization ambiguity, which was recently fixed by different
techniques \cite{DJS01,I04,BDF04}.
These additional steps 
allowed the compilation of following 3PN accurate expressions for
the orbital energy and angular momentum, in harmonic gauge 
and in the center-of-mass frame, reduced by $\mu$ as
\bs
\bea
{ E} &=&
 {{ E}}_{0} + 
\frac{1}{c^2} \, {{ E}}_{1} 
+ \frac{1}{c^4} \, {{E}}_{2} 
+ \frac{1}{c^6} \,{{ E}}_{3}\,, 
\\
J &=& \left| {\bf R} \times {\bf V} \right|
\biggl \{ { J}_{0} + 
\frac{1}{c^2} \,{ J}_{1} 
+ \frac{1}{c^4} \,{ J}_{2} 
+ \frac{1}{c^6} \,{ J}_{3}   \biggr \}\,,
\eea
\label{H_E_Ja}
\es    
where ${\bf R}$ and ${\bf V}$ are the relative separation and velocity vectors.
The various contributions to $E$ and $J$ 
are given by
\bs
\bea
{{E}}_{0} & = & \frac{1}{2}v^{2}-\frac{1}{r},
\label{E0}\\
{{ E}}_{1} & = & 
\frac{3}{8}(1-3\eta)v^{4}
+\frac{1}{2}(3+\eta)\frac{v^{2}}{r}
+\frac{1}{2}\eta\frac{\dot{r}^{2}}{r}+\frac{1}{2}{\left (\frac{1}{r}\right )}^{2}
 \,,
\label{E1} \\
{{ E}}_{2} & = &\frac{5}{16}(1-7\eta+13\eta^{2})v^{6}+\frac{1}{8}(21-23\eta-27\eta^{2})\frac{v^{4}}{r}
\nonumber\\
 & & 
+\frac{1}{4}\eta (1-15\eta)\frac{v^{2}\dot{r}^{2}}{r}
-\frac{3}{8}\eta (1-3\eta)\frac{\dot{r}^{4}}{r} +\frac{1}{8}(14-55\eta
\nonumber\\
 & & 
+4\eta^{2}){\left(\frac{v}{r}\right)}^{2}
+\frac{1}{8}(4+69\eta
+12\eta^{2}){\left (\frac{\dot{r}}{r}\right )}^{2}
\nonumber\\
 & & 
-\frac{1}{4}(2+15\eta){\left (\frac{1}{r}\right )}^{3}\,,
\label{E2}
\\
{{E}}_{3}&=& \frac{1}{128}\bigg (35-413\eta+1666\eta^{2}-2261\eta^{3}\bigg )v^{8}
 +\frac{1}{16}\bigg(55-215\eta
 \nonumber\\
 & & 
 +116\eta^{2}+325\eta^{3}\bigg)\frac{v^{6}}{r}
 +\frac{1}{16}\eta(5-25\eta+25\eta^{2})\frac{\dot{r}^{6}}{r}
    \nonumber\\
 & & 
 -\frac{1}{16}\eta(21+75\eta-375\eta^{2})\frac{v^{4}\dot{r}^{2}}{r}
 -\frac{1}{16}\eta(9-84\eta
   \nonumber\\
 & & 
   +165\eta^{2})\frac{v^{2}\dot{r}^{4}}{r}
 +\frac{1}{16}\bigg(135-194\eta+406\eta^{2}
    -108\eta^{3}\bigg){\left (\frac{v^{2}}{r}\right )}^{2} 
    \nonumber\\
 & & 
 +\frac{1}{16}\bigg(12
  +248\eta-815\eta^{2}
    -324\eta^{3}\bigg){\left (\frac{v\dot{r}}{r}\right )}^{2} 
     \nonumber\\
 & & 
   -\frac{1}{48}\eta(731-492\eta-288\eta^{2}){\left (\frac{\dot{r}^{2}}{r}\right )}^{2}
   \nonumber\\
 & &  
 +\frac{1}{2240}\bigg [2800-\left (53976-1435\pi^{2}\right )\eta-11760\eta^{2} 
  \nonumber\\
 & & 
 +1120\eta^{3}\bigg ]{\frac{v^{2}}{r^{3}}}
 +\frac{1}{2240}\bigg [3360
  +\left (18568-4305\pi^{2}\right )\eta 
  \nonumber\\
 & & 
 +28560\eta^{2} +7840\eta^{3}\bigg ]{\frac{\dot{r}^{2}}{r^{3}}} 
 + \frac{1}{840}(315 + 18469\eta){\left (\frac{1}{r}\right )}^{4}\,,
  \\
{J}_{0} &=&1 \,,\\
{J}_{1} &=&\frac{1}{2}(1-3\eta)v^{2}+(3+\eta)\frac{1}{r} \,,
\label{J1} \\
{J}_{2} &=&\frac{3}{8}(1-7\eta+13\eta^{2})v^{4}
  +\frac{1}{2}(7-10\eta-9\eta^{2})\frac{v^{2}}{r} 
   \nonumber\\
 & &
 -\frac{1}{2}\eta(2+5\eta)\frac{\dot{r}^{2}}{r} 
 + \frac{1}{4}(14-41\eta+4\eta^{2}){\left (\frac{1}{r}\right )}^{2}
\,,\\
 {J}_{3}&=&
  +\frac{1}{16}(5-59\eta+238\eta^{2}-323\eta^{3} )v^{6}+\frac{1}{8}\bigg(33-142\eta+106\eta^{2}
  \nonumber\\
 & & 
  +195\eta^{3}\bigg)\frac{v^{4}}{r} 
 -\frac{1}{4} \eta (12-7\eta-75\eta^2)\frac{v^{2}\dot{r}^{2}}{r}
 +\frac{3}{8}\eta(2-2\eta
 \nonumber\\
 & & 
 -11\eta^{2})\frac{\dot{r}^{4}}{r}
+\frac{1}{12}\bigg(135-322\eta+315\eta^{2} 
-108\eta^{3}\bigg){\left (\frac{v}{r}\right )}^{2}
\nonumber\\
&&
+\frac{1}{24}\bigg (12-287\eta-951\eta^{2}
 -324\eta^{3}\bigg){\left (\frac{\dot{r}}{r}\right )}^{2}
 \nonumber\\
&&
+
\left [ \frac{5}{2}-\frac{1}{1120}\left (20796-1435\pi^{2}\right )\eta-7\eta^{2}+\eta^{3}\right ]{\left (\frac{1}{r}\right )}^{3} 
\,.
\label{J3}
\eea
\label{H_E_Jb}
\es
In the above expressions, $r = | {\bf r}|$ 
with ${\bf r}={\bf R}/(GM)$,
$v^2 = {\bf v} \cdot {\bf v}$ and $\dot r = \frac{ {\bf v} \cdot {\bf r}}{r}$,
where ${\bf v} = \frac{d {\bf r}}{dt}$ and $t$ is the coordinate time
scaled by $GM$.
The above expressions also match with those presented in \cite{MW04}.  

Since the orbital motion is restricted to a plane, 
we express the components of
${\bf r}$ 
in polar coordinates as
$ {\bf r} = r ( \cos \phi, \sin \phi) $ and obtain
${\bf v} =  ( \dot r\,\cos \phi
-r\,\dot \phi \,\sin \phi, \dot r\,
\sin \phi + r\,\dot \phi\, \cos \phi)
$ allowing us
to write $ v^2 = \dot r^2 + r^2\,\dot \phi^2$.
Using these inputs and 3PN accurate expressions for ${E}$ and $J$, given by
Eqs.\ (\ref{H_E_Ja}) and (\ref{H_E_Jb}), 
we obtain, after lengthy iterations, 
3PN accurate expressions for 
${\dot r}^2$ and $\dot \phi$,
in terms of $E, h=\frac{J}{G M}, \eta$ and 
$r$.
This leads to 3PN accurate expressions
for ${\dot r}^2$ and $(d\phi/ds)$, in harmonic coordinates, 
displayed in the appendix \ref{AppA}.
We note that the expressions for ${\dot r}^2$ and $\dot \phi$
differ from similar ones derived for ADM-type gauge at 
2PN and 3PN orders.
However, note that in both gauges the expressions for $\dot r^2$ and 
$\dot \phi $ are polynomials of degree seven.
It is interesting to note that in the previous section the orbital 
dynamics was fully determined with the help of the Hamiltonian.
However, here we employed neither the Hamiltonian nor the Lagrangian 
to determine the orbital dynamics. Though the energy is numerically 
equal to the Hamiltonian, we required both the energy and angular
momentum as functions of particle positions and velocities 
to determine the binary dynamics.
The analogy to energy as thermodynamical potential versus caloric and
thermal equations of state is quite appropriate \cite{Pathria}.

Following exactly the same procedure, detailed in the 
previous section,
we derive the 3PN accurate
orbital parametrization in harmonic gauge.
The third post-Newtonian accurate generalized quasi-Keplerian parametrization,
in harmonic coordinates, for compact binaries 
moving in  eccentric orbits  
is given by
\bs
\bea
r &=& a_r \left ( 1 -e_r\,\cos u \right )\,,\\
l \equiv n \left ( t - t_0 \right ) &=& 
u -e_t\,\sin u + 
\left ( \frac{g_{4t}}{c^4} + \frac{g_{6t}}{c^6} \right )\,(v -u)
\no
&&
 + 
\left ( \frac{f_{4t}}{c^4} + \frac{f_{6t}}{c^6} \right )\,\sin v
+ \frac{i_{6t}}{c^6} \, \sin 2\, v 
+ \frac{h_{6t}}{c^6} \, \sin 3\, v \,,\\
\frac{ 2\,\pi}{ \Phi} \left (\phi - \phi_{0} \right )
&=& v + 
\left ( \frac{f_{4\phi}}{c^4} + \frac{f_{6\phi}}{c^6} \right )\,\sin 2v
+
\left ( \frac{g_{4\phi}}{c^4} + \frac{g_{6\phi}}{c^6} \right )\, \sin 3v
\no
&&
+ \frac{i_{6\phi}}{c^6}\, \sin 4v
+ \frac{h_{6\phi}}{c^6}\, \sin 5v \,,
\eea
\es                       
where $ v = 2 \arctan \biggl [ \biggl ( \frac{ 1 + e_{\phi}}{ 1 - e_{\phi}}
\biggr )^{1/2} \, \tan \frac{u}{2} \biggr ]$.
The explicit 3PN accurate expressions for the orbital elements 
and functions of the
generalized quasi-Keplerian parametrization, in harmonic coordinates,
read
\bs 
\bea
{\it a_r}&=& \frac{1}{{(-2E)}} \bigg\{ 1+{\frac {{(-2E)}}{4\,{c}^{2}} \left( -7+\eta \right) }
+\frac{{{(-2E)}}^{2}}{16  {c}^{4}} \bigg[1+ {\eta}^{2}
\no&&
+{\frac{16}{{{(-2E {h}^{2})}}}( {-4+7\,
\eta}}) \bigg]
+{\frac {{{{(-2E)}}}^{3}}{6720\, {c}^{6}}}\, \bigg[ 105
-105\,\eta\no&&+105\,{\eta}^{3}
+{
\frac{1}{{{(-2E{h}^
{2})}}}} \bigg( 26880+4305\,{\pi }^{2}\eta-215408\,\eta
\no&&
+47040\,{\eta}^{2} \bigg)
-{\frac {4}{{{(-2E{h}^{2})}}^{2}}}\bigg(53760-176024\,\eta+4305\,{\pi }^{2}\eta
\no&&
+15120\,{\eta}^{
2} \bigg) 
\bigg] \bigg\},
\\
{e_{{r}}}^{2}&=&1+{{2E{h}^{2}}}+
{\frac {{(-2E)}}{4\,{c}^{2}}} \bigg\{ 24-4
\,\eta+5\, \left(-3+ \eta \right) {{(-2E{h}^{2})}} \bigg\}
\no&&
+
\frac{{{(-2E)}}^{2}}{8{c}^{4}}  \bigg\{ 60+148\,\eta+2\,{\eta}^{2} - {{(-2E{h}^{2})}}\left( 
80-45\,\eta+4\,{\eta}^{2} \right)
\no&&
+\,{\frac {32}{{{(-2E{
h}^{2})}}}(4-7\,\eta)}\bigg\}                       
+{\frac {{{(-2E)}}^{3}}{6720\,{c}^{6}}}
\bigg\{ -3360
+181264\,\eta
\no&&
+8610\,{\pi }^{2}\eta-67200\,{\eta}^{2}
+105{{(-2E{h}^{2})}}
\, \bigg(-1488+1120\,\eta
\no&&
-195\,{\eta}^{2}+ 4\,{\eta}^{3}
 \bigg)  -{\frac {80\,}{{{(-2E{h}^{2})}}}}\bigg(1008-21130\,\eta+861\,{\pi 
}^{2}\eta
\no&&
+2268\,{\eta}^{2}\bigg)
+{\frac {16\,}{{{(-2E{h}^{2})}}^{2}}\bigg(53760
-176024\,\eta+4305\,{\pi }^{2}\eta}
\no&&
+15120\,{\eta}^{2}\bigg) \bigg\}, 
\\
{\it n}&=&{{(-2E)}}^{3/2} \bigg\{ 1
+
{\frac {{{(-2E)}} }{8\,{c}^{2}}}\,(-15+\eta )
+{\frac {{{(-2E)}}^{2}}{128\,{c}^{4}}}\, \bigg[ 555+
30\,\eta
\no&&
+11\,{\eta}^{2}+{\frac {192\,}{\sqrt {{{(-2E{h}^{2})}}
}}(-5+2\,\eta)} \bigg] 
+{\frac {{{(-2E)}}^{3}}{3072\, {c}^{6}}}
\bigg[ -29385
\no&&
-4995\,
\eta-315\,{\eta}^{2}+135\,{\eta}^{3}+{\frac {5760}{\sqrt {{(-2Eh^2)}}}(17-9\,\eta+2
\,{\eta}^{2}})
\no&&
-\,{\frac {16}{{{(-2E{h}^{2})}}^{3/2}}\bigg(10080-13952\,\eta+123\,{\pi 
}^{2}\eta+1440\,{\eta}^{2}}\bigg) \bigg] \bigg\}, 
\\
{e_{{t}}}^{2}&=&1+2E{h}^{2}+
{\frac {{(-2E)}}{4\,{c}^{2}}}\bigg\{ -8+8
\,\eta- {{(-2E{h}^{2})}}( -17+7\,\eta)  \bigg\}
\no && 
+\frac{{{(-2E)}}^{2}}{8\, {c}^{4}} \bigg\{12+ 72\,\eta+20\,{\eta}^{2}-24 \sqrt {{(-2E h^2)}}
\left( 
-5+2\,\eta \right)
\no&&
- {{(-2E{h}^{2})}}( 112-47\,\eta+16\,{\eta}^{2}
)
-{\frac {16}{{{(-2E{h}^{2})}}}(-4+7\,\eta)}
\no&&
+{\frac {24}{\sqrt {{(-2E{h}^{2})}}}(-5+2\,\eta)}\bigg\}
+{\frac 
{{{(-2E)}}^{3}}{6720\, {c}^{6}}} \bigg\{ 23520-464800\,\eta
\no&&
+179760\,{\eta}^{2}+16800\,{
\eta}^{3}
-2520\, \sqrt {{(-2Eh^2)}} (265 -193\,\eta
\no&&
+46\,{\eta}^{2})-525{{(-2E{h}^{2})}}\bigg( -528+200\,\eta 
-77\,{\eta}^{2}
+24
\,{\eta}^{3}\bigg)
\no&&
-{\frac {6}{{{(-2E{h}^{2})}}}\bigg(73920-260272\,\eta
+4305\,{\pi }^{2}\eta+61040\,{\eta}^{2}}\bigg)
\no&&
+{
\frac {70}{
\sqrt {{(-2Eh^2)}}}\bigg(16380-19964\,\eta+123\,{\pi }^{2}\eta+3240\,{\eta}^{2}}\bigg)
\no&&
+{\frac {8}{{{(-2E{h}^{2})}}^{2}}\bigg(53760-176024\,\eta+4305\,{\pi }^{2}\eta+15120\,{\eta}^{2}\bigg)
}
\no&&
-{\frac {70}{{{(-2E{h}^{2})}}^{3
/2}}\bigg(10080-13952\,
\eta+123\,{\pi }^{2}\eta+1440\,{\eta}^{2}}\bigg) \bigg\},
\\
g_{{4\,t}}&=&
-\frac{3{{(-2E)}}^{2}}{2}\,
\bigg\{{\frac {1}{\sqrt {{(-2Eh^2)}}}}(-5+2\,\eta)\bigg\},
\\
g_{{6\,t}}&=&{\frac {{{(-2E)}}^{3}}{192}}\,
\bigg\{
{\frac {1}{{{(-2E{h}^{2})}}^{3/2}}}\bigg(10080-13952\,\eta+123\,{\pi }^{2}\eta
\no&&
+1440\,{\eta}^{2}\bigg)
+{\frac {1}{\sqrt 
{{(-2E{h}^{2})}}}(-3420+1980\,\eta-648\,{\eta}^{2}})
\bigg\},
\\                               
f_{{4\,t}}&=&-\frac{{{(-2E)}}^{2}}{8}\,
\bigg\{
{\frac {\sqrt {1+2E{h}^{2}}}{
\sqrt {{(-2Eh^2)}}}\eta\, \left( -15+\eta \right) }
\bigg\},
\\
f_{{6\,t}}&=&{\frac {{{(-2E)}}^{3}}{2240}}\,\bigg\{
\frac {1}{{{(-2E{h}^{2})}}^{3/2}\sqrt {1+2E{h}^{2}}}\bigg(22400+43651\,\eta
\no&&
-1435\,{
\pi }^{2}\eta
-20965\,{\eta}^{2}+385\,{\eta}^{3}\bigg)
\no&&
+{
\frac {1}{\sqrt {{(-2E{h}^{2})}}\sqrt {1+2E{h}^{2}}}}\bigg(-22400-
49321\,\eta
\no&&
+27300\,{\eta}^{2}+1435\,{\pi }^{2}\eta-1225\,{\eta}^{3}\bigg)
\no&&
+{
\frac {35\,\sqrt {{{(-2E{h}^{2})}}}}{\sqrt {1+2E{h}^{2}}}} \, \eta\, (\,297 
-175\,\eta+23\,{\eta}^{2}) 
\bigg\},
\\
i_{{6\,t}}&=&\frac{{{(-2E)}}^{3}} {16}\bigg\{
\,{\frac { 1+2E{h}^{2} }{{{(-2E{h}
^{2})}}^{3/2}}}\,\eta\left(116-49\,\eta+3\,{\eta}^{2} \right)  \bigg\},
\\
h_{{6\,t}}&=&{\frac {{{(-2E)}}^{3}}{192}}\,\bigg\{
{\left(\frac{1+2E{h}^{2}}{{{(-2E{h}^{2})}}}\right)^{3/2}}\eta\, \left( 23-73\,\eta+13\,{\eta}^{2} \right) 
\bigg\},
\\
\Phi&=&2\,\pi \, \bigg\{ 1+{\frac {3}{{h}^{2}{c}^{2}}}+
\frac{{{(-2E)}}^{2}}{4\,{c}^{4}}
\bigg[ {\frac {3}{{{(-2E{h}^{2})}}}}(-5+2\,\eta)
\no&&
-{\frac {15}{{{(-2E{h}^{2})}}^{2}}}(-7+
2\,\eta)
\bigg]
+{\frac {{{(-2E)}}^{3}}{128\,{c}^{6}}}\,
 \bigg[{\frac { 5}{{{(-2E{h}^{2})}}^{3}}}\bigg(7392
 \no&&
 -8000\,\eta+336\,{\eta}^{2}+123\,{\pi }^{2}\eta
\bigg)+{\frac {24}{{{(-2E{h}^{2})}}}}(5-5\,\eta+4\,{\eta}^{2})
\no&&
-{\frac {1}{{{(-2E{h}^{2})}}^{2}}}\bigg(10080-13952\,\eta+123\,{\pi }^{2}\eta+
1440\,{\eta}^{2}\bigg) \bigg] \bigg\}, 
\\
f_{{4\,\phi}}&=&
\frac{{{(-2E)}}^{2}}{8}
\bigg\{
{\frac {1+
{{2E{h}^{2}}} }{{{(-2E{h}^{2})}}^{2}}} \left( 1+19\,\eta-3\,{\eta}^{2} \right)
\bigg\},  
\\
f_{{6\,\phi}}&=&{\frac {{{(-2E)}}^{3}}{26880}}\bigg\{
{\frac {1}{{{(-2E{h}^{2})}}^{3}}}\bigg(67200+994704\,\eta-30135\,{\pi }^{2}
\eta
\no&&
-335160\,{\eta}^{2}-4200\,{\eta}^{3}\bigg)
+{\frac {1}{{{{(-2E{h}^{2})}}}^{2}}}\bigg(-60480-991904\,\eta
\no&&
+30135\,{\pi }^{2}\eta+
428400\,{\eta}^{2}-8400\,{\eta}^{3}\bigg)
+{\frac {1}{{{(-2E{h}^{2})}}}}\bigg(840
\no&&
+141680\,
\eta-99960\,{\eta}^{2}+10080\,{\eta}^{3}\bigg)
\bigg\},
\\
g_{{4\,\phi}}&=&-\frac{{{(-2E)}}^{2}}{32}\bigg\{
{\frac { \left( 1+{{2E{h}^{2}}} \right) ^{3/2}}{{{(-2E{h}^{2})}}^{2}}}\, \eta\, \left( -1+3\,
\eta \right)                                                                       
\bigg\},
\\
g_{{6\,\phi}}&=&
{\frac {{{(-2E)}}^{3}}{{8960}}\,\frac{1}{\sqrt {1+2E{h}^{2}}}}
\bigg\{-35\,\eta\, ( 14-49
\,\eta+26\,{\eta}^{2} ) 
\no&&
-{\frac {1}{{{(-2E{h}^{2})}}}}\,\eta \bigg(-36196+
1435\,{\pi }^{2}+ 29225\,\eta-2660\,{\eta}^{2} \bigg) 
\no&&
+{\frac {1}{{{(-2E{h}^{2})}}^{2}
}}\,\eta \bigg(-71867+2870\,{\pi }^{2}+ 56035\,
\eta-2275\,{\eta}^{2} \bigg)
\no&&
-{\frac {1}{{{(-2E{h}^{2})}}^{3}}}\, \eta \bigg(-36161 +1435\,{\pi }^{2}+28525\,\eta-525\,{\eta}^{2}
 \bigg) 
\bigg\},
\\
i_{{6\,\phi}}&=&{\frac {{{(-2E)}}^{3}}{192}}\,\bigg\{
{\frac {( 1+2E{h}^{2})^{2}}{{{(-2E{h}^{2})}}^{3}}}\,\eta\, ( 82-57\,\eta+15\,{\eta}^{2})
\bigg\},
\\
h_{{6\,\phi}}&=&{\frac {{{(-2E)}}^{3}}{256}}\,
\bigg\{
{\frac {( 1+2E{h}^{2})^{5/2}}{{{(-2E{h}^{2})}}^{3}}}\, \eta\, (1-5\,\eta+5\,{\eta}^{2})
\bigg\},
\\
{e_{{\phi}}}^{2}&=&1+2E{h}^{2}+
{\frac {{{(-2E)}}}{4\,{c}^{2}}}\, \bigg\{ 24+ {{(-2E{h}^{2} )}}( -15+\eta ) \bigg\}
\no&&
+
\frac{{{(-2E)}}^{2}}{16 {c}^{4}}\,\bigg\{ -40+34\,\eta +18\,{\eta}^{2}
-{{(-2E{h}^{2})}} (160 
\no&&
-31\,
\eta+3\,{\eta}^{2} ) -{\frac {1}{{{(-2E{h}^{2})}}}}(-416+91\,\eta+15\,{
\eta}^{2}) \bigg\} 
\no&&
+
{\frac {{{{(-2E)}}}^{3}}{13440\, {c}^{6}}}\,\bigg\{ -584640-17482\,\eta
-4305\,{\pi }^{2}\eta-7350\,{\eta}^{2}
\no&&
+8190\,{\eta}^{3}-420 {{(-2E{h}^{2})}}\bigg( 744-248\,\eta+31\,{\eta}^{2}
+3\,{
\eta}^{3}\bigg)
\no&&
-{\frac {14}{{{(-2E{h}^{2})}}
}}\bigg(36960-341012\,\eta+4305\,{\pi }^{2}\eta-225\,{\eta}^{2}
\no&&
+150\,{\eta}^{3}\bigg)
-{\frac {1}{{{(-2E{h}^{2})}}^{2}}}\bigg(-2956800+5627206\,\eta
\no&&
-81795\,{\pi }^{2}\eta+14490\,{\eta}^{2}+7350
\,{\eta}^{3}\bigg) \bigg\}.
\eea
\es 
In harmonic coordinates too, there are PN accurate relations connecting 
the three eccentricities $e_r, e_t$ and $e_\phi$. These relations
read
\bs
\bea
{ e_t}&=&{ e_r}
\bigg\{
 1+  {\frac {{ (-2E)}}{2{c}^
{2}}}(3\eta-8)+ {\frac {{{ (-2E)}}^{2}}{4{c}^
{4}}}    {\frac {1}{{ (-2E{h}^{2})}}}
 \bigg[
 -16+28\eta
 \no&&
 + (-30+12\eta)\sqrt {{ (-2Eh^2)}}
 +(36-19\eta+ 6{\eta}^{2} ) { (-2E{h}^{2})}\bigg]
 \no&&
+
{\frac {{{ (-2E)}}^{3}}{6720{c}^{6}}}    {\frac {1
}{{h}^{4}{{ (-2E)}}^{2}}}
\bigg[
-215040+704096\eta-17220{\pi}^{2}\eta
\no&&
-60480{\eta}^{2}
+
35\bigg( -10080+13952\eta-123{\pi}^{2}\eta
\no&&
-1440{\eta}^{2} \bigg) \sqrt {{ (-2Eh^2)}}
+
\bigg( 87360-354848\eta+4305{\pi}^{2}\eta
\no&&
+105840{\eta}^{2} \bigg){ (-2E {h}^{2})}
+
\bigg(-134400+54600\eta-28560{\eta}^{2}
\no&&
+8400{\eta}^{3} \bigg) {{ (-2E{h}^{2})}}^{2}
+
(321300 -225540\eta
\no&&
+52920{\eta}^{2}) {{ (-2E{h}^{2})}}^{3/2}
\bigg]
\bigg\}\,,\\
{ e_\phi}&=&{ e_r}
\bigg\{
1+  {\frac {{ (-2E)}}{2{c}^
{2}}}  \eta+  {\frac {{{ (-2E)}}^{2}}{32{c}^{4}}
}    {\frac {1}{{ (-2E{h}^{2})}}}
\bigg[ 
160+357\eta-15{\eta}^{2}
\no &&
+(-\eta+ 11{\eta}^{2} ) { (-2E{h}^{2})}
\bigg]
+
{\frac {{{ (-2E)}}^{3}}{8960{c}^{6}}}
{\frac {1}{{{ (-2E{h}^{2})}}^{2}}}
\bigg[
412160
\no&&
+1854\eta-18655{\pi}^{2}\eta-166110{\eta}^{2}-2450{\eta}^{3}
+ \bigg( 24640 
\no&&
-182730\eta+7175{\pi}^{2}\eta+156520{\eta}^{2}-5250{\eta}^{3} \bigg){ (-2E {h}^{2})} 
\no&&
+ 70 \eta (-1-{\eta}+31{\eta}^{2}) {{(-2E{h}^{2})}}^{2}
\bigg]
\bigg\}\,.
\eea
\es

  We observe that the structure of the 3PN accurate parametrization is
  identical in both ADM-type and harmonic coordinates. This is not surprising
as the expressions for
$\dot r^2$
 and $\dot \phi$ are polynomials of same degree 
in $s = \frac{1}{r}$
in these two gauges.
We also observe that since the equations of motion,
and hence the expressions for $\dot r^2$ and $\dot \phi$,
are the same in ADM and harmonic gauges to 1PN order, the 
orbital elements and the relations between the
eccentricities to 1PN order are also identical
in these two gauges.
It is also possible to connect the radial eccentricities, and hence 
other eccentricities, associated with parametrizations
in ADM-like and harmonic coordinates.
This indicates that a circular orbit in ADM-type gauge also implies 
a circular orbit in harmonic gauge.
The relation connecting $e_r^2$ in ADM-lik coordinates to that in 
harmonic coordinates reads
\bea 
e_r^2|_{\rm H}
&=& 
e_r^2|_{\rm A}
\bigg\{ 1
+\frac{{{ (-2E)}}^{2}}{4{c}^{4}}\bigg[
5\eta+\frac {2}{{ E{h}^{2}}}(17\eta+1) \bigg ]
\no&&
+\frac{{{ (-2E)}}^{3}}{1680{c}^{6}}
\bigg[3570\eta-630{\eta}^{2}
+{\frac{1}{ (-2E{h}^{2})}}\bigg(420-23520{\eta}^{2}
\no&&
+
({72844} -2205{\pi }^{2})
\eta\bigg)+{\frac{4} {{{
(-2E{h}^{2})}}^{2}}}\bigg(-2520+8400{\eta}^{2}
\no&&
-(58004
-2205{\pi }^{2} ) \eta\bigg)\bigg] \bigg\}\,,
\eea
where  $e_r^2|_{\rm H} $ and $e_r^2|_{\rm A}$ are the expressions 
$e_r^2$ in harmonic and ADM-type coordinates.

          The most striking result is that the expressions for 
$n$ and $\Phi$ in terms of $E$ and $h$
are identical to 3PN order in both ADM-type and harmonic 
coordinates. Indeed it was shown,                                               using the Hamilton-Jacobi approach to 
describe the relative 
trajectory of the binary, that the functional form of
 $n$ and $\Phi$ should be independent 
of the coordinate system used,
if expressed in terms of gauge (coordinate) invariant
quantities like $E$ and $h$ \cite{DS88}. 
This should be contrasted with the functional forms for
other orbital elements like $a_r$ and $e_r$,
in terms of $E$ and $h$, which depend on the 
coordinate system used. 
The explicit verification of the above prediction constitutes a powerful
check on the algebra involved in the derivation of 
3PN accurate orbital representation in these two gauges.

  The gauge invariance of $n$ and $\Phi$ allows us to obtain 
following gauge invariant expressions for the third post-Newtonian accurate adimensional
orbital energy and angular momentum for compact binaries of arbitrary
mass ratio, moving in eccentric orbits. 
\bs
\bea 
\frac{E}{c^2} &=&
-\frac{x}{2} \biggl \{  1+ \frac{1}{12}\, \biggl [ 15-\eta \biggr ]\, x
+\frac{1}{24}\, \biggl [ \biggl ( 120 -48\,\eta \biggr)
\left( {\frac{k'}{x}} \right )^{1/2} + 15 -15\,\eta-{\eta}^{2} 
\biggr ] 
{x}^{2}
\no
&&
+{\frac {1}{5184}}\, \biggl [  
\biggl ( 68040
+ \left( -173376+2214\,{\pi }^{2} 
\right) \eta 
+12960\,{\eta}^{2}
\biggr)\,
\left ( \frac{k'}{x} \right )^{3/2}
\no
&&
+ \biggl ( 68040
-30240\,\eta 
-6048\,{\eta}^{2}
\biggr)\,
\left ( {\frac{k'}{x}} \right )^{1/2} 
-4995
-6075\,\eta
\no
&&
-450\,{\eta}^{2} 
-35\,{\eta}^{3}
\biggr ]\,{x}^{3} \biggr \}\,,\\ 
(h\,c)^2 &=& \frac{1}{k'}
\biggl \{ 1
+\frac{1}{4}\,\biggl [ \left( 35-10\,\eta \right) \,
\left ( \frac{k'}{x} \right ) 
-5 +2\,\eta \biggr ]\, x
+{\frac {1}{384}}\, \biggl [ \biggl ( 7560
+ \left( 615\,{\pi }^{2}-23200 \right) \eta 
\no
&&
-720\,{\eta}^{2 }
\biggr )
\left ( \frac{k'}{x} \right )^{2}
+ \biggl ( -5880
+ \left( 11072
-123\,{\pi }^{2}
 \right) \eta 
-960\,{\eta}^{2}
\biggr ) 
\left ( \frac{k'}{x} \right ) 
\no
&&
- 480
+160\,\eta
 +80\,{\eta}^{2}
\biggr ]\, x^2  
 \biggr \} \,,
\eea
\label{Eq.27}
\es
where $x = \left( \frac{G\,M\,n}{c^3} \right)^{2/3}$ and 
$k'= \left ( \frac{ \Phi - 2\,\pi}{6\,\pi} \right )$.
The above equations generalize to eccentric orbits 
the gauge invariant expression connecting the orbital energy of
a compact binary in circular orbit to its period at 3PN order
\cite{BFIJ02,BDF04}.
Using 3PN accurate far-zone fluxes, which are not yet computed, and
Eqs.~(\ref{Eq.27}),
it should be possible to compute 3PN accurate expressions for 
$ dx/dt $ and $ dk'/dt$ in terms of $x$ and $k'$.
This would generalize the 3PN accurate 
gauge invariant expression for the rate of change of the orbital
frequency of a compact binary in circular orbit, available in
\cite{BFIJ02,BDFI}.
The Eqs.~(\ref{Eq.27}) also indicate that it should be possible to 
characterize non-circular orbits in post-Newtonian relativity
in an invariant manner.  Therefore, this parametrization
should be useful 
to analyze existing general relativistic simulations 
involving compact binaries in a post-Newtonian framework.

\section{The Summary and discussions}
\label{Disc}
In this paper, we have presented the third post-Newtonian accurate 
`Keplerian' type parametrization for the motion of two non-spinning
compact objects moving in an eccentric orbit.
The above 3PN accurate parametrization, which is structurally quite similar to 
the 2PN accurate generalized quasi-Keplerian
representation, is given in both ADM-type and harmonic coordinates.
The associated orbital elements and functions were explicitly computed 
in terms of the 3PN accurate conserved orbital energy, angular momentum 
and the finite mass ratio. 
We explicitly showed that to even these high post-Newtonian orders
there are gauge invariant quantities to characterize
an eccentric orbit in post-Newtonian relativity.
We also performed, for the first time, some clever consistency checks 
to validate the lengthy algebraic and trigonometric
manipulations involved in the derivation of the parametrization.

  There are quite a few possible applications for our parametrization and some 
of them are currently  under investigation.
The representation will be required to construct `ready to use' search templates
for inspiraling eccentric compact binaries, whose orbital dynamics is fully
3.5PN accurate, extending the currently available 2.5PN accurate ones,
presented in \cite{DGI}.
Recently, using a solution to 3PN accurate equations of motion
for compact binaries moving in non-circular orbits,  gauge dependent
expressions for the associated conserved orbital 
energy and angular momentum were obtained 
which were used to analyze general relativistic simulations 
involving compact binaries \cite{MW02,MW04}.
Using the arguments that there are gauge 
invariant expressions to  3PN order in our parametrization 
and that the orbital elements of the representation, when radiation reaction 
is included, are continuously evolving variables,
it should be possible to construct a better 
post-Newtonian accurate `diagnostic tool'
to analyze the orbital configuration and 
motion of inspiralling compact binaries.

Finally, we note that to obtain ${\cal O}(1/c^{11})$ corrections to the
orbital averaged expressions for the
far-zone fluxes associated with inspiralling eccentric 
binaries, 3PN corrections to the quadrapole approximation,
our parametrization in harmonic coordinates will have to be heavily 
employed. The other crucial ingredients required for those 
computations, namely the 
${\cal O}(1/c^{11})$ corrections to the 
far-zone fluxes for compact binaries
in general orbits, in harmonic coordinates, are expected to be available in the 
near future  \cite{BDFI}.

\begin{acknowledgments}                                                                 
We thank T. Oikonomou for compiling and checking 
some inputs. 
The financial support of
the Deutsche
Forschungsgemeinschaft (DFG) through SFB/TR7
``Gravitationswellenastronomie'' is gratefully acknowledged.

\end{acknowledgments}

\appendix
\section{The equations governing the 3PN accurate radial and angular motion
in ADM-type and harmonic coordinates}
\label{AppA} 
In this appendix, we display the 3PN accurate expressions for $\dot r^2=\frac{1}{s^4}\left(\frac{d s}{dt}\right)^2$
and $ \frac{d \phi}{ds}=\dot \phi/\frac{d s}{dt}$ that
are essential to obtain  
the generalized quasi-Keplerian parametrization in ADM-type and 
harmonic coordinates
in terms of $E, h, \eta$ and $s$. 
As mentioned earlier, the
expression for $\dot r^2=\frac{1}{s^4}\left(\frac{d s}{dt}\right)^2$ is required to obtain
3PN accurate turning points of the radial motion and 
hence required in the computations that
determine the 3PN accurate radial parametrization
and `Kepler equation'.
We first exhibit the 3PN accurate expression for $\dot r^2$
 in ADM-type coordinates.
\bs 
\bea
\dot r ^2 \equiv  \frac{1}{s^4}
{\left({\frac {{ ds}}{{ dt}}}\right)}^{2}&=& a_{{0}}+a_{{1}}s+
a_{{2}}{s}^{2}+a_{{3}}{s}^{3}+a_{{4}}{s}^{4}+a_{{5}}{s}^{5}+a_{{6}}{s}
^{6}+a_{{7}}{s}^{7} \,,
\eea
\bea
a_{{0}}&=&2E +{\frac {3{E}^{2} }{{c}^{2}}} (-1+3\eta) +{\frac {{E}^{3}}{{c}^{4}}} (4-19\eta+16{\eta}^{2
})
 \no
&&
 +{\frac {{E}^{4}}{4{c}^{6}}} \bigg(-20
+128\eta-211
{\eta}^{2}+56{\eta}^{3}\bigg)\,, 
\\
a_{{1}}&=&2+{\frac {2E }{{c}^{2}}} (-6+7\eta)+
{\frac {6{E}^{2}}{{c}^{4}}  (3-16\eta+7{\eta}^{2}) 
}
\no &&
+{\frac {2{E}^{3} }{{c}^{6}}} \bigg(-12
+93\eta-171{\eta}^{2}+35{\eta}^{
3}\bigg)\,,
\\
a_{{2}}&=&-{h}^{2} +{\frac{1}{{c}^{2}}} \bigg[5(-2+\eta)+2{h}^{2}E(1-3
\eta) \bigg]
\no&&+{\frac {1}{{c}^{4}}}\bigg[E(37
-122\eta
+36{\eta}^{2})
+3{E}^{2}{h}^{2}
 (-1+5\eta
 -5{\eta}^{2}) \bigg] 
  \no&&
+{\frac {1}{2{c}^{6}}}\bigg[{E}^{2} \bigg( -111
+
1034\eta         
-1268{\eta}^{2}
+228{\eta}^{3}\bigg) 
\no&&
+2{E}^{3}{h}^{2}
 \bigg( 4-27\eta+50{\eta}^{2}-20{\eta}^{3} \bigg)\bigg]\,, 
 \\
a_{{3}}&=&{\frac {{h}^{2}}{{c}^{2}}}(8-3\eta)
+{\frac{1}{2{c}^{4}}}  \bigg[ (53-83
\eta
+20{\eta}^{2})
\no&&
+2
{h}^{2}E(-16+73\eta-19{\eta}^{2}) 
\bigg]  
+
{\frac {1}{24{c}^{6}}}
\bigg[
{E} \bigg(-
1896-( 3{\pi}^{2}
\no&&
-14252 ) \eta-10848{\eta}^{2}+1776
{\eta}^{3}\bigg) +
12{E}^{2}{h}^{2}\bigg(48 
-348\eta
\no&& 
+603{\eta}^{2}-96{\eta}^{3}\bigg) 
\bigg]\,,
\\
a_{{4}}&=&{\frac {3{h}^{2}}{{c}^{4}}} (-11+11\eta-2{\eta}^{2})
 +{\frac {1}{48{c}^{6}}}
 \bigg[
 \bigg(-2400
+ ( 57{\pi}^{2}
 \no&&
+9524 ) \eta-5292{\eta}^{2}+768{\eta}^{3}\bigg)
 +
 12{h}^{2}E \bigg(264
 -1857\eta
 \no&&
 +1206{\eta}^{2}
 -120{\eta}^{3}\bigg)
 \bigg]\,,
\\
a_{{5}}&=&-{\frac {{h}^{4}}{4{c}^{4}}} (4\eta+{\eta}^{2}) +\frac{1}{96 {c}^{6}}
\bigg[ 
 {h}^{2} \bigg( 9024
 + \left( -
21908+3{\pi}^{2} \right) \eta
\no&&
+7872{\eta}^{2}
-192{\eta}^{3}
 \bigg) +24E{h}^{4}(12\eta-34{\eta}^{2}-27{\eta}^{3}) \bigg]\,,
 \no
a_{{6}}&=&{\frac{{h}^{4}}{4{c}^{6}}} \bigg( 71\eta+2{\eta}^{2}-32{\eta}
^{3} \bigg)\,,
\no
a_{{7}}&=&{\frac {13{h}^{6}}{8{c}^{6}} {\eta}^{3} }\,,
\eea  
\es
To obtain the post-Newtonian accurate `Kepler equation' and to parameterize
the angular motion, we usually employ the PN accurate expressions for $\frac{ds}{dt}$
factorized by the PN accurate roots $s_+$ and $s_-$.

Instead of displaying the original expression for $\frac{d \phi}{d s} = \dot \phi / \frac{d s}{d t}$,
we exhibit below the expression for $\frac{d \phi}{d s}$ factorized by the PN accurate
roots $s_+$ and $s_-$ which was actually employed in the computations.
\bs
\bea
\frac{d\phi}{ds}&=&-\frac{B_0+{B_1}s+{B_2}s^2+B_3 s^3
+B_4s^4+B_5 s^5}{\sqrt{(s_--s)(s-s_+)}}\,,
\eea
where PN accurate functions $B_i$ are given by
\bea
B_{{0}}&=&1+\frac {1}{2{h}^{2}{c}^{2}}(6-\eta)
+\frac{1}{{c}^{4}}\bigg[
\frac{1}{8{h}^{4}}{{(176-80\eta+3{\eta}^{2})}}
+\frac{E}{2{h}^{2}}(15
\no&&
-12\eta+2{\eta}^{2})
\bigg]
+  
\frac{1}{{c}^{6}}
\bigg[
\frac{1}{16{h}^{6}}\bigg(3376+ \left( 40{\pi}^{2}-3784 \right) \eta
\no&&
+350{\eta}^{2}-5{\eta}^{3}\bigg)
+\frac{E}{48{h}^{4}}
\bigg(6120+ \left( 57{\pi}^{2}-9592 \right) \eta
\no&&
+1884{\eta}^{
2}
-72{\eta}^{3}\bigg)
+\frac{{E}^{2}}{4{h}^{2}}\bigg(15-19\eta+32{\eta}^{2
}-6{\eta}^{3}\bigg)\bigg]\,,
\\
B_{{1}}&=&{\frac {1}{2{c}^{2}}}\eta
+\frac{1}{{c}^{4}}\bigg[\frac{1}{4{h}^{2}}(17-5\eta-{
\eta}^{2})+\frac{E}{4}(2\eta+5{\eta}^{2})\bigg]
\no&&
+
 {\frac {1}{{c}^{6}}}
\bigg[
\frac{1}{16{h}^{4}}\bigg(740+ \left( 20{\pi}^{2}-1114 \right) \eta+6{\eta}^{2}+3{
\eta}^{3}\bigg)
\no&&
+{\frac {E}{96{h}^{2}}}\bigg(864+ \left( -1708-3{
\pi}^{2} \right) \eta+480{\eta}^{2}-12{\eta}^{3}\bigg)
\no&&
+\frac{{E}^{2}}{8}
(5{\eta}^{2}-{\eta}^{3})\bigg]\,,
\\
B_{{2}}&=&\frac{1}{8{{c}^{4}}}(4\eta+17{\eta}^{2})
+
\frac{1}{{c}^{6}}
\bigg[\frac{1}{16{h}^{2}}
\bigg(64+ \left( 10{\pi}^{2}-338 \right) \eta+86{\eta}^{2}
\no&&
-17{
\eta}^{3}\bigg)
+\frac{E}{8}(-33\eta-28{\eta}^{2}+45{\eta}^{3})\bigg]\,,
\\
B_{{3}}&=&-\frac{9{h}^{2} }{8{c}^{
4}}{\eta}^{2} 
+
\frac{1}{c^6}
\bigg[{\frac {1}{64}}
\bigg( -100+3{\pi}^{2}) \eta
-568{\eta}^{2}+504{\eta}^{3}\bigg)
\no&&
-4{h}^{2}E{\eta}^{3}
\bigg]\,,
\\
B_{{4}}&=&\frac{{h}^{2}}{16{c}^{6}}( 20\eta+112{\eta}^{2}-121{
\eta}^{3})\,,
 \\
B_{{5}}&=&\frac{15{h}^{4}}{16{c}^
{6}}{\eta}^{3}\,.
\eea
\es

In harmonic coordinates, the explicit expressions for  $\dot r^2$
and $\dot{\phi}$ are extracted from
the 3PN accurate expressions for conserved 
orbital energy and angular momentum,
in the center of mass frame, given by Eqs.~(\ref{H_E_Ja}) and 
(\ref{H_E_Jb}).
To 3PN order,  $\dot r^2$
 in harmonic coordinates reads
\bs 
\bea
\dot r ^2 \equiv  \frac{1}{s^4}
{\left({\frac {{ ds}}{{ dt}}}\right)}^{2}&=& a_{{0}}+a_{{1}}s+
a_{{2}}{s}^{2}+a_{{3}}{s}^{3}+a_{{4}}{s}^{4}+a_{{5}}{s}^{5}+a_{{6}}{s}
^{6}+a_{{7}}{s}^{7} \,,
\eea
\bea
a_{{0}}&=&2E+{\frac {3{E}^{2}}{{c}^{2}}}(-1+3\eta)
+
\frac { {E}^{3}}{{c}^{4}} (4-19\eta+16{\eta}^{2})
\no&&
+
\frac {{E}^{4}}{4{c}^{6}}
(-20+128\eta-211{\eta}^{2}+56{\eta}^{3})\,,
\\
a_{{1}}&=& 2+{\frac {2E}{{c}^{2}}}(-6+7\eta)+{\frac {6{E}^{2}}{{c}^{4}}}(3-
16\eta+7{\eta}^{2})
\no&&
+{\frac {2{E}^{3}}{{c}^{6}}}(-12+93\eta-
171{\eta}^{2}+35{\eta}^{3})\,,
\\
a_{{2}}&=&-{h}^{2}+{\frac {1}{{c}^{2}}}\bigg[5(-2+\eta)+2{h}^{2
} (1-3\eta) E\bigg]
\no&&
+{\frac {1}{{c}^{4}}}\bigg[ E (36-127
\eta+36{\eta}^{2}) +3{h}^{2}{E}^{2} (-1+5\eta-5{\eta}
^{2}) \bigg]
\no&&
+
{\frac {1}{{c}^{6}}}\bigg[\frac{{E}^{2}}{6}(-324+3476\eta-4017
{\eta}^{2}+684{\eta}^{3})
\no&&
+{h}^{2} {E}^{3} (4-27\eta+50{
\eta}^{2}-20{\eta}^{3}) \bigg]\,,
\\
a_{{3}}&=&{\frac {{h}^{2}}{{c}^{2}}}(8-3\eta) +{
\frac {1}{{c}^{4}}}\bigg[\frac{1}{2}(52-81\eta+20{\eta}^{2})+{h}^{2} E(-16
\no&&
+61\eta
-19{\eta}^{2})\bigg] 
+
\frac{1}{c^6}
\bigg[{\frac {E}{840}}\bigg(-63840+
 ( 661412
 \no&&
 +4305{\pi }^{2} ) \eta-400260{\eta}^{2}+62160
{\eta}^{3}\bigg) 
\no&&
+\frac{{h}^{2}{E}^{2}}{2}\bigg(48-288\eta+399{\eta}^{2}-96{
\eta}^{3}\bigg) \bigg]\,,
\\
a_{{4}}&=&
{\frac {{h}^{2} }{4{c}^{4}}}(-132+75\eta-24{\eta}^{2})
+
\frac{1}{c^6}\bigg[{\frac {1}{840}}\bigg(-42000
\no&&
+\left( 241364+4305
{\pi }^{2} \right) \eta-93450{\eta}^{2}+13440{\eta}^{3}\bigg)  
\no&&
+\frac{{h}^{2} E }{4}
\bigg(264-1887\eta+552{\eta}^{2}-120{\eta}^{3}\bigg) \bigg]\,,
\\
a_{{5}}&=&\frac{{h}^{4}}{4{c}^{4}}(15\eta-{\eta}^{2}) 
+
\frac{1}{{c}^{6}}
\bigg[{\frac {{h}^{2}}{1120}} \bigg(107520+ ( -1435{
\pi }^{2}
\no&&
-259344 ) \eta+22960{\eta}^{2}-2240{\eta}^{3}\bigg)
  \no&&
  +\frac{{h}^{4} E}{4}(-71\eta+241{\eta}^{2}-27{\eta}^{3}) \bigg]\,,
\\
a_{{6}}&=&{\frac {{h}^{4}}{8{c}^{6}}} (-11\eta+358{\eta}^{2}-64{
\eta}^{3})\,,
\\
a_{{7}}&=&{\frac {{h}^{6} }{8{c}^{6}}}(23\eta-73{\eta}^{2}+13{\eta}^{3})\,.
\eea           
\es 
As in ADM-type gauge, the 3PN accurate
factorized expression for $\frac{d \phi}{d s}$ reads  
\bs
\bea
\frac{d\phi}{ds}&=&-\frac{B_0+{B_1}s+{B_2}s^2+B_3 s^3
+B_4s^4+B_5 s^5}{\sqrt{(s_--s)(s-s_+)}}\,.
\eea
The explicit expressions for the $B_i$ in harmonic coordinates are
\bea
B_{{0}}&=&1+\frac{1}{2{h}^{2}{c}^{2}}{(6-\eta)}
+ 
\frac{1}{{c}^{4}}
\bigg[\frac{1}{8{h}^{4}}(172-148\eta+3{\eta}^{2})+\frac{E}{4{h}^{2}}
(28
\no&&
-63\eta+4{\eta}^{2})\bigg]
+
\frac{1}{{c}^{6}} 
\bigg[{\frac {1}{1680{h}^{6}}}
\bigg(341880+ ( 8610{\pi }^{2}
\no&&
-640588 ) \eta+74970{\eta}^{2}-525{\eta}^{3}\bigg)
+{\frac {E}{840{h}^{4}}}\bigg(99960
\no&&
+ \left( 4305{
\pi }^{2}-345446 \right) \eta+76125{\eta}^{2}-1260{\eta}^{3}\bigg)
\no&&
+\frac{{E}^{2}}{24{h}^{2}} \bigg(84
-1055\eta+567{\eta}^{2}-36{\eta}^{3}\bigg)\bigg]\,,
\\
B_{{1}}&=&\frac{1}{2{c}^{2}}\eta
+
\frac{1}{{c}^{4}}
\bigg[\frac{1}{4{h}^{2}}(16-22\eta-{\eta}^{2})
+\frac{5E}{4}(\eta+{\eta}^{2})\bigg]
\no&&
+       
\frac{1}{{c}^{6}}
\bigg[{\frac {
1}{1680{h}^{4}}}\bigg(73920+ \left( -196604+4305{\pi }^{2} \right) 
\eta+5460{\eta}^{2}
\no&&
+315{\eta}^{3}\bigg)
+{\frac {E}{3360{h}^{2}}}
\bigg(26880+ \left( -191048+4305{\pi }^{2} \right) \eta
\no&&
+31080{
\eta}^{2}-420{\eta}^{3}\bigg)
+\frac{{E}^{2}}{8}(-9\eta+28{\eta}^{2}-{
\eta}^{3})\bigg]\,,
\\
B_{{2}}&=&\frac{1}{8{c}^{4}}(4+67\eta+17{\eta}^{2})
+
\frac{1}{{c}^{6}}
\bigg[{\frac 
{1}{3360{h}^{2}}}\bigg(18480+ ( 4305{\pi }^{2}
\no&&
-8444) \eta-
1050{\eta}^{2}-3570{\eta}^{3}\bigg)
+\frac{E}{24}(134\eta+411{
\eta}^{2}
\no&&
+135{\eta}^{3})\bigg]\,,
\\
B_{{3}}&=&\frac{{h}^{2} }{8{c}^{4}}(3\eta-9{\eta}^{2})
+\frac{1}{c^6}\bigg[{\frac {1}{6720}}\bigg( \left( -56296-12915{\pi }^{2}
 \right) \eta  
 \no&&
 +103320{\eta}^{2}+52920{\eta}^{3}\bigg)+\frac{{h}^{2}E}{4}
 (\eta+4{\eta}^{2}-16{\eta}^{3}) \bigg]\,,
 \\
B_{{4}}&=&\frac{{h}^{2} }{48{c}^{6}}(581\eta-45{\eta}^{2}-363{
\eta}^{3})\,,
\\
B_{{5}}&=&\frac{5{h}^{4}}{16{c}^{6}}(\eta-5{\eta}^{2}+5{
\eta}^{3})\,.
\eea
 \es 
We note that the expressions for $\dot r^2$
and $\frac{d \phi}{ds}$ in these two gauges have the same structure though the
coefficients differ at 2PN and 3PN orders.

\section{A sketch of some computational details} 
\label{AppB}
This appendix details the computation required to obtain 
the 3PN accurate `Kepler equation' and the
parametrization for the angular motion.
As explained in Sec.~\ref{3PN_ADM}, the temporary form for the 3PN accurate
$l(u)$ relation, given by Eq.~(\ref{temp_t}), follows from 
Eqs.~(\ref{3PN_t_t0}) and (\ref{3PN_T}) by employing the auxiliary
variable $\tilde v$.
This computation is not straightforward as the direct evaluation of 
the integral for $(t-t_0)$, given by Eq.~(\ref{temp_t}), 
gives a PN accurate expression in terms of $E, h, \eta, s_{-},s_{+}$
and $s$.
To obtain the temporary form for the 3PN accurate `Kepler equation', we
multiply the above result with the 3PN accurate expression for $n$,
and use the
trigonometric relations given below
\bs
\bea
s& =& \frac{1}{a_r\,(1-e_r\cos(u))} =
\frac{1+e_r \cos \tilde v}{a(1-e_r^2)}\,,\\
u &=& \arccos \left( \frac {s_-+s_+}{s_--s_+}-2\frac{s_-s_+}
{(s_--s_+)s} \right) \,,\\
\tilde v &=& 
\arccos \left(\frac {2s}{s_--s_+}-\frac{s_-+s_+}{s_--s_+}\right)
\,.
\eea
\es
These relations are also employed heavily to obtain the temporary
parametrization for the angular motion, given by Eq.~(\ref{temp_phi}).

 Let us turn our attention to the details of the computation,
 which gave the final parametrization for $\frac{2\,\pi}{\Phi}
\, \left ( \phi - \phi_0 \right )$.
The starting points of the above calculation  are Eq.~(\ref{temp_phi})
and the introduction of PN accurate true anomaly 
$ v = 2 \arctan \biggl [ \biggl ( \frac{ 1 + e_{\phi}}{ 1 - e_{\phi}}
\biggr )^{1/2} \, \tan \frac{u}{2} \biggr ]$,
where $e_{\phi}$ differs from $e_r$ by yet to be determined PN corrections.
It is easy to obtain the following 3PN 
accurate expression for $\tilde v$ in terms of $v$, which reads
\bea 
\tilde v&=&v-\frac{y}{c^2} \sin v
 + \frac{{y}^{2}}{4c^4}\left(\sin 2v +2\sin v  \right)
 \no&&
 \label{tildevofv}
- \frac{{y}^{3}}{12c^6}\left( 3\sin v  
+3\sin 2v +\sin 3v \right) 
\eea
where $y$ is expressible in terms of those 
PN corrections 
which relate $e_{\phi}$ and $e_r$.
Using the above relation, we  express $\frac{2\,\pi}{\Phi}
\, \left ( \phi - \phi_0 \right )$, given by Eq.~\ref{temp_phi},
in terms of $v$ and demand that there are no `$\sin v$' terms to 3PN 
order. This requirement, as explained in Sec.~\ref{3PN_ADM}, is motivated 
by the desire that we want for the first post-Newtonian order `Keplerian' like 
parametrization for the angular part.
This procedure uniquely defines PN corrections connecting $e_{\phi}$ to $e_r$
and leads us to the final parametrization for the angular
motion, given by Eq.~(\ref{fin_phi}).

  Finally, we rewrite the temporary parametrization for the 
$l(u)$ relation, given by Eq.~(\ref{temp_t}), in terms of $v$
which leads us to the final expression for 3PN accurate `Kepler
equation' as given by Eq.~(\ref{K_Eqn_3PN}).

\newpage 

\begin{thebibliography}{999}


\bibitem{S91}
B. F. Schutz, in {\it The Detection of Gravitational Waves},
edited by D. G. Blair, Cambridge: Cambridge University Press (1987),
p. 330.

\bibitem{2PN_hcp}
L. Blanchet, T. Damour, B. R. Iyer,
C. M. Will, and A. G. Wiseman, Phys.\ Rev.\ Lett.\ {\bf 74}, 3515
(1995); 
L. Blanchet, T. Damour, and B. R. Iyer,  Phys.\
Rev.\ D {\bf 51}, 5360 (1995); 
C. M. Will and A. G. Wiseman, Phys.\ Rev.\ D {\bf 54}, 4813 (1996);
L. Blanchet, B. R. Iyer, C. M. Will, and A. G. Wiseman,
 Class.\ Quantum Grav.\ {\bf 13}
575 (1996).

\bibitem{CFPS93}
C. Cutler, L. S. Finn, E. Poisson, and G. J. Sussmann, 
Phys.\ Rev.\ D {\bf 47}, 1151 (1993); T.\ Damour, B.\ R.\ Iyer, and 
B.\ S.\ Sathyaprakash, Phys.\ Rev.\ D {\bf 57}, 885 (1998).

\bibitem{BDFI}
L. Blanchet, T. Damour, G. Esposito-Far\`ese, and B.R. Iyer,
{\it Gravitational radiation from inspiralling compact binaries completed at the
third post-Newtonian order}, gr-qc/0406012.

\bibitem{DGI}
T. Damour, A. Gopakumar, and B. R. Iyer,
{\it Phasing of gravitational waves from inspiralling eccentric binaries},
gr-qc/0404128.

\bibitem{NR_rev}
T.\ W.\ Baumgarte and S.\ L.\ Shapiro,
Phys.\ Rept.\  {\bf 376}, 41 (2003).


\bibitem{MW02}
T. Mora and C. M. Will,
Phys.\ Rev.\ D {\bf 66}, 101501(R) (2002).

\bibitem{MW04}
T. Mora and C. M. Will,
Phys.\ Rev.\ D {\bf 69}, 104021 (2004).

\bibitem{DD86}
T. Damour and N. Deruelle, Ann. Inst. Henri Poincar\'e
Phys. Theor. {\bf 44}, 263 (1986).

\bibitem{DT92}
T. Damour and J. Taylor, Phys. Rev. D {\bf  45},  1840 (1992).

\bibitem{DD85}
T. Damour and N. Deruelle, Ann. Inst. Henri Poincar\'e
Phys. Theor. {\bf 43}, 107 (1985).



\bibitem{IS03_LR}
I.\ H.\ Stairs, Living Rev.\  Relativ.\  {\bf 6}, 5 (2003).

\bibitem{K51004}
M. Kramer, A. G. Lyne, M. Burgay, A. Possenti, R. N. Manchester, F. Camilo,
M. A. McLaughlin, D. R. Lorimer, N. D'Amico, B. C. Joshi, J. Reynolds,
and P. C. C. Freire, {\it The double pulsar -- A new testbed
for relativistic gravity}, in ``Binary Pulsars'' Eds.\ F.\ Rasio
and I.\ Stairs (2004),
astro-ph/0405179.

\bibitem{DS88}
T. Damour and G. Sch\"afer, Nuovo Cimento B
{\bf 101}, 127 (1988).


\bibitem{SW93}
G. Sch\"afer and N. Wex, Phys. Lett. {\bf 174 A},
196, (1993); erratum {\bf 177}, 461.

\bibitem{ME01}
S.\ Menzel, {\it Untersuchung zur letzten stabilen Kreisbahn zweier Schwarzer
L\"ocher}, ``diploma thesis'',  Friedrich Schiller University Jena (2001), unpublished.

\bibitem{CM_TB}  D. Brouwer and G. M. Clemence,
{\it Methods of Celestial Mechanics}, Academic Press (1961).

\bibitem{ADM62}
R. Arnowitt, S. Deser, and C. W. Misner,
{\it The Dynamics of General Relativity}
in ``Gravitation: An Introduction to Current Research'', edited by L. Witten, 
Wiley (1962), gr-qc/0405109.

\bibitem{OOKH74}
T. Ohta, H. Okamura, T. Kimura, and K. Hiida, Prog. Theor. Phys. {\bf 51}, 1598 (1974).

\bibitem{DS85}
T. Damour and G. Sch\"afer, Gen. Rel. Grav. {\bf 17}, 879 (1985).


\bibitem{JS98}
P. Jaranowski and G. Sch\"afer, Phys.\ Rev.\ D {\bf 57}, 7274 (1998),\\
P. Jaranowski and G. Sch\"afer, Phys.\ Rev.\ D {\bf 60}, 124003 (1999).

\bibitem{DJS00a}
T. Damour, P. Jaranowski, and G. Sch\"afer, 
Phys. Rev. D {\bf 62}, 021501 (2000).

\bibitem{DJS01}
T. Damour, P. Jaranowski, and G. Sch\"afer, Phys. Lett. B {\bf 513}, 147 (2001).

\bibitem{DJS00b}
T. Damour, P. Jaranowski, and G. Sch\"afer, 
Phys. Rev. D {\bf 62}, 044024 (2000).

\bibitem{BI03}
L. Blanchet and B. R. Iyer, 
Class.\ Quantum Grav.\ {\bf 20},
755 (2003).

\bibitem{LB_lr}
L.\ Blanchet, Living Rev.\  Relativ.\  {\bf 5}, 3 (2002).
 
\bibitem{BF00}
L. Blanchet and G. Faye, Phys. Lett. A {\bf 271}, 58 (2000).

\bibitem{BF01}
 L. Blanchet and G. Faye, Phys. Rev. D {\bf 63}, 062005 (2001).

\bibitem{IF03}
 Y. Itoh and T. Futamase, Phys. Rev. D {\bf 68}, 121501 (2003).

\bibitem{I04}
Y. Itoh, Phys. Rev. D {\bf 69}, 064018 (2004).

\bibitem{BDF04}
L. Blanchet, T. Damour and G. Esposito-Far\`ese, 
Phys.\ Rev.\ D {\bf 69}, 124007 (2004).

\bibitem{Pathria}
R.K. Pathria,
{\em Statistical Mechanics}, Butterworth-Heinemann (1996).

\bibitem{BFIJ02}
L. Blanchet, G. Faye, B.\ R.\ Iyer, and B.\ Joguet,
Phys.\ Rev.\ D {\bf 65}, 061501 (2002).


\end{thebibliography}
\end{document}